\documentclass{article}
\usepackage{PRIMEarxiv}
\errorcontextlines\maxdimen

\usepackage[utf8]{inputenc} 
\usepackage[T1]{fontenc}    
\usepackage{url}            
\usepackage{booktabs}       
\usepackage{nicefrac}       
\usepackage{microtype}      
\usepackage{fancyhdr}       
\usepackage{graphicx}
\usepackage{xcolor}
\usepackage{threeparttable}
\usepackage{booktabs}
\usepackage{multirow}
\usepackage{tabularx}
\usepackage{arydshln}
\usepackage{subcaption}

\usepackage{lineno,hyperref}
\hypersetup{colorlinks, citecolor=blue, linkcolor=red, urlcolor=green}
\usepackage{amsmath,amsfonts,amssymb}
\usepackage{graphics}
\usepackage{bm}
\usepackage{multirow}
\usepackage{algorithm}
\usepackage{algpseudocode}

\usepackage{mathtools}
\usepackage{siunitx}
\DeclarePairedDelimiterXPP\BigOSI[2]%
  {\mathcal{O}}{(}{)}{}%
  {\SI{#1}{#2}}

\title{Stochastic Smoothed Particle Hydrodynamics for Stochastic Mechanics Problems}

\author{
Mridul Tiwari \\
Department of Applied Mechanics \\
Indian Institute of Technology Delhi \\
New Delhi, 110016, India \\
\texttt{am1210975@am.iitd.ac.in} \\
\And
Sawan Kumar \\
Department of Applied Mechanics \\
Indian Institute of Technology Delhi \\
New Delhi, 110016, India \\
\texttt{amz228095@am.iitd.ac.in} \\
\And
Md Rushdie Ibne Islam \thanks{Corresponding author}\\
Department of Ocean Engineering and Naval Architecture \\
Indian Institute of Technology Kharagpur \\
West Bengal 721302, India \\
\texttt{mrii@naval.iitkgp.ac.in} \\
\texttt{https://rushdieislam.github.io/} \\
\And
Souvik Chakraborty \\
Department of Applied Mechanics \\
Indian Institute of Technology Delhi \\
New Delhi, 110016, India \\
Yardi School of Artificial Intelligence (ScAI) \\
Indian Institute of Technology Delhi \\
New Delhi, 110016, India \\
\texttt{souvik@am.iitd.ac.in} \\
\texttt{https://www.csccm.in/}
}

\begin{document}
\maketitle
\begin{abstract}
Smoothed Particle Hydrodynamics (SPH) is a mesh‑free Lagrangian method renowned for modeling large deformations and free‑surface flows, yet classical formulations remain confined to deterministic systems. We introduce Stochastic SPH (S‑SPH), which employs orthogonal Polynomial Chaos expansions to represent uncertainties in system parameters, forcing functions, and boundary or initial conditions, while spatial variation is captured via the SPH kernel. Random fields are discretized through Karhunen-Loève expansions, and a Galerkin projection in the polynomial basis transforms the underlying SPDE into a coupled system of ordinary differential equations governing the time evolution of expansion coefficients. To enforce Dirichlet and Neumann conditions in a mesh‑free context, ghost‑particle techniques augmented by a gradient‑correction matrix are employed, and a predictor–corrector integration scheme ensures numerical stability. We validate S‑SPH on benchmark problems, including one‑dimensional advection with stochastic advection speed, inviscid Burgers’ equations with random initial amplitudes, and two‑dimensional Burgers’ flows with uncertain Fourier‑mode initial fields and viscosity, demonstrating excellent agreement with Monte Carlo simulation statistics of mean and variance. Remarkably, S‑SPH achieves up to three orders of magnitude reduction in computational cost relative to direct sampling approaches. The proposed framework thus provides an efficient, accurate, and fully mesh‑free methodology for uncertainty quantification in complex mechanics applications.
\end{abstract}


\section{Introduction}
The modeling of complex fluid and solid dynamics has become crucial in a wide array of fields, from astrophysics \cite{sph_exa2_2024,spmhd_2023,Springel2010,Violeau2012} and geophysics \cite{mullet2023sph,ono2008sph} to engineering disciplines \cite{rogers2020sph,feng2023sph,Price2012,LiuMeshfree2003}. Smoothed Particle Hydrodynamics (SPH) has emerged as a powerful and versatile mesh-free approach, particularly well-suited for simulating problems involving large deformations, free-surface flows, and complex interface phenomena. This method, which was initially developed by Gingold and Monaghan in the late 1970s \cite{Gingold1977} and later advanced by Monaghan in the 1990s \cite{Monaghan1992}, is widely recognized for its ability to model fluid flows and large material deformations without relying on a fixed grid or background mesh. The fundamental idea is to represent fluids and solids as collections of discrete particles, each corresponding to a material point, with the state variables and their spatial derivatives approximated using kernel functions that essentially approximate a Dirac-delta function. These particles interact with one another based on physical laws, such as the conservation of mass, momentum, and energy. The kernel-based approximation of the state variables allows SPH to model complex phenomena such as high-velocity impact and penetration \cite{SHAW20093962,Johnson1996SPHFH,MEHRA2006318}, fragmentation \cite{mullet2023caldera,CHEN20242215}, dynamic crack propagation \cite{Liu2023RockFracture, Feng2024PDSPHCoupling}, and fluid-structure interaction \cite{Ono2023FSIFramework, Antoci2007SPHFEM} that are often difficult to model with traditional grid-based methods.

Over the years, considerable efforts have been dedicated to improving the SPH method, addressing several challenges that initially limited its effectiveness and broadened its range of applications. For example, the field variable approximations near the boundary suffer from $C^0$ and $C^1$ inconsistencies i.e., even the constant or linear functions cannot be represented accurately. The treatments of free surfaces or the interfaces between different media in the simulation domain are also challenging. Over time, various consistency correction algorithms and boundary treatment techniques have been developed, including the use of ghost particles \cite{Vela2019GhostPositioning,Anonymous2011BoundaryIntegralSPH,Colagrossi2017ShockSPH}, reflective boundary conditions \cite{Anonymous2024Reflective3D,Schechter2012GhostSPH,Anonymous2024ImprovedDBC}, and specialized algorithms for handling solid-fluid interactions \cite{Liu2020FSIValidation,Geni2018SPHCoupling,Anonymous2021SPHDEM}. These advances ensure that the simulation more accurately reflects physical boundaries, which is crucial for simulations involving complex geometries or interactions between solids and fluids. Similarly, different forms of SPH have been developed to improve spatial resolution and the fidelity of the simulations. Some other key contributions in SPH include the development of Weakly Compressible SPH \cite{Shadloo2011WCSPH}, Incompressible SPH \cite{Solenthaler2009PCISPH} most commonly used in fluid flow simulations, Total Lagrangian SPH \cite{Febrianto2024TLSPH} most commonly used in the modelling of solid deformation. In parallel with these advances, computational techniques have been enhanced to make SPH simulations more efficient and scalable. SPH is embarrassingly parallel in nature, and the developments of more sophisticated algorithms, such as parallel computing techniques and GPU acceleration \cite{Krog2010GPUSPH,Crespo2011GPU,Harada2007SPHGPU}, have allowed SPH to handle larger and more complex systems, making it practical for real-world applications that require high computational power. These improvements in efficiency have made SPH more reliable for large-scale simulations, enabling the modelling of phenomena that were previously computationally prohibitive.

Despite the effectiveness of SPH in solving a wide array of problems, as highlighted above, existing SPH algorithms are only tailored towards deterministic systems governed by deterministic partial differential equations. In real-world applications, factors such as measurement errors, model simplifications, or inherent variability in material properties and external conditions can introduce uncertainties. To ensure safety and reliability, it is important to consider these uncertainties into the governing physics and study its influence on the predicted response variables. One potential alternative for modeling uncertainty is to employ Monte Carlo Simulation (MCS) \cite{10.5555/539488} in conjunction with numerical solvers such as finite element method or SPH. However, MCS based approaches requires large number of deterministic simulations and are computationally expensive. Improvements to classical MCS such as Latin Hypercube Sampling \cite{SEAHOLM198897}, stochastic collocation method \cite{Babuska2007StochasticCollocation}, and Importance Sampling \cite{Glynn1989ImportanceMarkov}  have been proposed; however, the computational cost continues to be prohibitive. Non-sampling-based approaches such as perturbation approach \cite{Zhang2020UncertaintyReview}  on the other hand are computationally efficient but only works for low-dimensional systems. Also, it necessitates the computation of the gradient of the output variable with respect to the random inputs, which is non-trivial to calculate. Non-intrusive surrogate model-based approaches, such as Polynomial Chaos Expansion \cite{Xiu2002}, Gaussian Process Regression \cite{Rasmussen2006GaussianProcesses}, Analysis of Variance decomposition \cite{Zhang2020UncertaintyReview}, etc., provide a trade-off between the MCS-based and non-sampling-based approaches. However, these approaches are data-driven in nature, and hence, it cannot be guaranteed that the underlying physics will be satisfied.

In theory, the effect of uncertainties on the state variables can be modeled as stochastic partial differential equations (SPDE) where the parameters, boundary conditions, forcing function, and/or initial conditions are represented as random variables. Unfortunately, solution of SPDE is non-trivial and in particular, current state-of-the-art SPH algorithms are not adept in handling SPDEs. To address this apparent gap, we here propose a stochastic SPH (S-SPH) algorithm for stochastic mechanics problems modeled as SPDEs. The proposed S-SPH algorithm exploits orthogonal polynomial \cite{sparsegrid,pce,multiscale} to model the uncertain variables and SPH kernel to capture the spatial variation. Time evolution is tracked by using the predictor-corrector algorithm \cite{mit_pc,dl_spde,sheffield_pc}. A key feature of the proposed approach resides in the fact that the response statistics can be analytically computed.

The remainder of this paper is organized as follows. Section~\ref{sec:problem} details the mathematical formulation and problem statement. In Section~\ref{sec:methods}, we describe the S-SPH framework in details along with the treatment of boundary conditions. Section~\ref{sec:numerical} presents several benchmark numerical examples to validate our approach. Finally, Section~\ref{sec:conclusion} provides the concluding remarks.

\section{Problem Statement}\label{sec:problem}

We define a spatial domain in $d$ dimensions, $\mathcal{X} \subset \mathbb{R}^{d}$ that is both open and bounded, along with a complete probability space $(\Omega, \mathcal{F}, \mathbb{P})$, where $\Omega$ represents the sample space, $\mathcal{F}$ is the associated $\sigma$-algebra, and $\mathbb{P}$ is the probability measure. To mathematically describe a dynamical system in a probabilistic framework that accounts for uncertainties, we model the system using a generalized SPDE as follows:
\begin{equation}\label{eq:spde}
(\bm{u}(\bm{x},t; \bm \xi))_t+\mathcal{L}_{b(\bm \xi)}\left(\bm{u}\right)\left(\bm x, t; \bm \xi\right) = \bm{f}(\bm{x},t;\bm{\xi}),
\end{equation}
where U $\subset \mathbb{R}^{d}$, the subspace consisting of the solution field $\bm{u}$, $\bm f : \mathbb{R}^{d+m_\xi+1} \mapsto \mathbb{R}^{d}$ is a function, $\mathcal{L} : U \subset \mathbb{R}^{d} \to \mathbb{R}^{d}$ denotes a differential operator with parameter $\bm b$, and $(\bm{u})_t$ represents the partial derivative of the field $\bm{u}$ with respect to time.
The vector $\bm{\xi} : \Omega \mapsto \Gamma \subset \mathbb{R}^{m_\xi}$ consists of independent random variables $\xi_1(\omega), \dots, \xi_{m_\xi}(\omega)$, where $\omega \in \Omega$. 
Note that Eq. \eqref{eq:spde} accounts for uncertainties originating due to randomness in the system parameters $\bm b(\bm \xi)$ or due to randomness in the forcing function $\bm f\left(\cdot ,\cdot; \bm \xi \right)$. We note that the system parameter $b\left(\bm \xi \right)$ can potentially vary with space and time; however, the same is not shown here for brevity of representation.

To solve the SPDE in Eq. \eqref{eq:spde}, additional constraints 
in form of boundary and initial conditions needs to be defined. Given the possible uncertainties in the boundary and initial condition, the same are defined in the spatiotemporal domain \([0, T] \times\mathcal{X}\) with boundary \(\partial \mathcal{X} = \partial \mathcal{X}_D \cup \partial \mathcal{X}_N\), where \(\partial \mathcal{X}_D\) and \(\partial \mathcal{X}_N\) correspond to the Dirichlet and Neumann boundaries, respectively.
\begin{equation}\label{eq:dir_bc}
    \bm{u}(\bm{x},t;\bm{\xi}) = \bm{g_1}(\bm{x},t;\bm{\xi}), \quad \text{on } \partial \mathcal{X}_D \times [0, T].
\end{equation}

\begin{equation}\label{eq:neu_bc}
    \frac{\partial \bm{u}(\bm{x},t;\bm{\xi})}{\partial \bm{n}} = \bm{g_2}(\bm{x},t;\bm{\xi}), \quad \text{on } \partial \mathcal{X}_N \times [0, T].
\end{equation}

\begin{equation}\label{eq:ic}
    \bm{u}(\bm{x},0;\bm{\xi}) = \bm{u}_0(\bm{x},\bm{\xi}), \quad \text{in } \mathcal{X}.
\end{equation}  
With this setup, the objective of this paper is to develop a novel SPH algorithm for solving generalized SPDE Eq. \eqref{eq:spde} given the boundary and initial condition in Eqs. \eqref{eq:dir_bc}--\eqref{eq:ic}.

\section{Stochastic smoothed particle hydrodynamics (S-SPH)}\label{sec:methods}
In this section, we discuss the proposed Stochastic SPH (S-SPH) including the functional decomposition to approximate the stochastic evolution of the field variables, Karhunen Lo\'{e}ve expansion for discretizing the input stochastic field, boundary correction, and gradient correction. We also provide a detailed algorithm to outlining the steps involved in the proposed approach.   

\subsection{Function approximation} 
Revisiting, Eq. \eqref{eq:spde}, the field variable $\bm u \left(\bm x, t;\bm \xi \right)$ can be represented using the Dirac delta function $\delta \left( \bm x - \bm x' \right)$ as,
\begin{equation}\label{eq:dirac_representation}
    \bm u (\bm x , t;\bm\xi) = \int_{\chi} u\left(\bm x',t;\bm \xi \right) \delta \left( \bm x - \bm x' \right)  d \bm x',
\end{equation}
where $\chi$ is the integral volume.
Similar to classical SPH, the Dirac delta function in S-SPH is approximated using the kernel function,
\begin{equation}\label{eq:s_sph}
    \bm u (\bm x , t;\bm\xi) \approx  \int_{\chi} u\left(\bm x',t;\bm \xi \right) W \left( \bm x - \bm x'; h \right)  d \bm x',
\end{equation}
where $W\left( \bm x - \bm x'; h \right)$ is the kernel function that is positive, radially symmetric, and normalized,
\begin{equation}\label{eq:kernel_norm}
    \int_\chi W(\bm x - \bm x';h)d\bm x' = 1.
\end{equation}
$h$ in Eqs. \eqref{eq:s_sph} and \eqref{eq:kernel_norm} is the smoothing length and dictates the influence radius.
In practice, the integral in Eq. \eqref{eq:s_sph} is further approximated as a summation over neighboring particles,
\begin{equation}\label{eq:s_sph_dis}
    {\bm u}\left( \bm x_i, t;\bm \xi\right) \approx \sum_{\bm x_j \in  N_{\bm x_i}}{\bm u \left( \bm x_j, t; \bm \xi \right) W \left( \bm x_i - \bm x_j; h \right)} \frac{m_{\bm x_j}}{\rho_{\bm x_j}},
\end{equation}
where $\bm x_i$ is the position of the $i-$th particle and $\bm x_j$ represents the position of the $j-$th particle while resides in the neighbor $N_{\bm x_i}$ of particle $\bm x_i$. $m_{\bm x_j}$ and $\rho_{\bm x_j}$ are the mass and density of the particle at position $\bm x_j$, respectively. Eq. \eqref{eq:s_sph_dis} effectively represents the field variable at position $\bm x_i$ based on the contribution from the neighboring particles.

We note that the field variable represented using Eq. \eqref{eq:s_sph_dis} depends on the stochastic variable $\bm x_i$. This renders direct application of the Eq. \eqref{eq:s_sph_dis} impractical. In response to this challenge, we propose to further decompose Eq. \eqref{eq:s_sph_dis} as follows,
\begin{equation}\label{eq:s_sph_ortho}
    {\bm u}\left( \bm x_i, t;\bm \xi\right) \approx \sum_{\bm x_j \in  N_{\bm x_i}}{ \underbrace{\left\{\sum_{\bm{\alpha}\in \mathcal{J}} \hat{\bm{u}}_{\bm{\alpha}}(\bm{x}_j, t) \, \Phi_{\bm{\alpha}}(\bm{\xi})\right\}}_{\bm u \left( \bm x_j, t; \bm \xi \right)}
     W \left( \bm x_i - \bm x_j; h \right)} \frac{m_{\bm x_j}}{\rho_{\bm x_j}},
\end{equation}
where $\hat{\bm u}_{\bm \alpha}$ represent the coefficients associated with the orthogonal polynomial $\Phi_{\bm \alpha}(\bm \xi)$. Note that in the orthogonal polynomial expansion in Eq. \eqref{eq:s_sph_ortho} enables us to separate that stochastic variables $\bm \xi$ from the spatial variables $\bm x$ and the temporal variable $t$. The multivariate orthogonal polynomial is further defined as
\begin{equation}\label{eq:multivariate}
\Phi_{\bm{\alpha}}(\bm{\xi}) = \prod_{i=1}^{p} \phi_{\alpha_i}(\xi_i),
\end{equation}
with \(\bm{\xi} = (\xi_1, \xi_2, \dots, \xi_p)\) being the vector of independent random variables, \(\bm{\alpha} = (\alpha_1, \alpha_2, \dots, \alpha_p) \in \mathbb{N}_0^p\) a multi-index and \(\phi_{\alpha_i}(\xi_i)\) is the univariate orthogonal polynomial of degree \(\alpha_i\) with respect to the probability measure of \(\xi_i\).
The index set \(\mathcal{J}\) is typically chosen as
\begin{equation}
\mathcal{J} = \left\{ \bm{\alpha} \in \mathbb{N}_0^p : |\bm{\alpha}| = \sum_{i=1}^p \alpha_i \leq q \right\},
\end{equation}
where \(q\) is the maximum degree of the expansion. 
\subsection{Computing the expansion coefficient}\label{subsec:coeff}
In the proposed S-SPH, we use Eq. \eqref{eq:s_sph_ortho} to represent the field variable. However, the unknown expansion coefficients are unknown and needs to be computed. To that end, we substitute Eq. \eqref{eq:s_sph_ortho} into the governing SDE in Eq. \eqref{eq:spde},
\begin{equation}\label{eq:spde_sph}
    \left(\sum_{\bm x_j \in  N_{\bm x_i}}{ \sum_{\bm{\alpha}\in \mathcal{J}} \hat{\bm{u}}_{\bm{\alpha}}(\bm{x}_j, t) \, \Phi_{\bm{\alpha}}(\bm{\xi})}
     W_{ij} V_{j} \right)_t + \mathcal L_{\bm b\left(\bm \xi \right)}\left(\sum_{\bm x_j \in  N_{\bm x_i}}{ \sum_{\bm{\alpha}\in \mathcal{J}} \hat{\bm{u}}_{\bm{\alpha}}(\bm{x}_j, t) \, \Phi_{\bm{\alpha}}(\bm{\xi})}
     W_{ij} V_{j} \right) = \bm f \left(\bm x_i,t; \bm \xi\right),
\end{equation}
where $W_{ij} = W(\bm x_i - \bm x_j; h)$ and $V_j = \frac{m_{\bm x_j}}{\rho _{\bm x_j}}$. Note that $\mathcal L _{\bm b \left(\bm \xi \right)}$ in Eqs. \eqref{eq:spde} and \eqref{eq:spde_sph} represents the generalized differential operator and can also be represented as a series expansion as follows,
\begin{equation}\label{eq:L_op}
\mathcal{L}_{\bm b(\bm \xi)} = \sum_{|\alpha| \leq m} a_\alpha(\bm{u};b) D^\alpha,
\end{equation}
where \( D^\alpha \) denotes the partial derivative operator, defined as $
D^\alpha = \frac{\partial^{|\alpha|}}{\partial x_1^{\alpha_1} \partial x_2^{\alpha_2} \cdots \partial x_n^{\alpha_n}}.$
The term \( |\alpha| = \alpha_1 + \alpha_2 + \cdots + \alpha_n \) represents the total order of the multi-index \( \alpha \), where  
$\alpha = (\alpha_1, \alpha_2, \ldots, \alpha_n), \quad \alpha_i \in \mathbb{N}_0$ is a multi-index with each \( \alpha_i \) being a non-negative integer. The expansion coefficients \( a_\alpha(\bm{u};\bm b) \) are dependent on the field function \( \bm{u} \) and the random parameter \( \bm b \), thereby introducing stochasticity into the differential operator. We note that for brevity of representation, the dependence of $\bm u$ on the spatial coordinate $\bm x$, temporal variable $t$, and stochastic variables $\bm \xi$ is not shown in Eq. \eqref{eq:L_op}. Similarly, the dependence of system parameter $b$ on the stochastic variable $\bm \xi$ is also not shown.

In S-SPH, the differential operator applied to the field variable can be represented as,
\begin{equation}\label{eq:s_sph_grad}
    D^\alpha\left( \bm u \right) \left(\bm x_i,t;\bm \xi \right) \approx \sum_{\bm x_j \in  N_{\bm x_i}}{\left[ {\left\{\underbrace{\sum_{\bm{\alpha}\in \mathcal{J}} \hat{\bm{u}}_{\bm{\alpha}}(\bm{x}_i, t) \, \Phi_{\bm{\alpha}}(\bm{\xi})}_{\bm u (\bm x_i,t;\bm \xi)} - \underbrace{\sum_{\bm{\alpha}\in \mathcal{J}} \hat{\bm{u}}_{\bm{\alpha}}(\bm{x}_j, t) \, \Phi_{\bm{\alpha}}(\bm{\xi})}_{\bm u (\bm x_j,t;\bm \xi)}\right\} }
    \underbrace{\left\{\frac{\partial^{|\alpha|}}{\partial x_1^{\alpha_1} \cdots \partial x_n^{\alpha_n}}W _{ij}\right\}}_{D^\alpha (W)\left(\bm x_i - \bm x_j\right)}
      \frac{m_{\bm x_j}}{\rho_{\bm x_j}}\right]}.
\end{equation}
Substituting Eqs. \eqref{eq:s_sph_grad} and \eqref{eq:L_op} into Eq. \eqref{eq:spde_sph}, we obtain
\begin{equation}\label{eq:spde_sph2}
    \left(\sum_{\bm x_j \in  N_{\bm x_i}}{ \sum_{\bm{\alpha}\in \mathcal{J}} \hat{\bm{u}}_{\bm{\alpha}}^j \, \Phi_{\bm{\alpha}}(\bm{\xi})}
     W_{ij} V_{j} \right)_t +  \sum_{|\alpha| \leq m} \sum_{\bm x_j \in  N_{\bm x_i}} \sum_{\bm{\alpha}\in \mathcal{J}} \left\{ a_\alpha(\cdot;b) { \hat{\bm{u}}_{\bm{\alpha}}^{ij}\, \Phi_{\bm{\alpha}}(\bm{\xi}) } 
    {D^\alpha W_{ij}}
      V_j\right\}
      = \bm f\left(\cdot,\cdot;\bm \xi\right),
\end{equation}
where $\hat{\bm{u}}_{\bm{\alpha}}^j = \hat{\bm{u}}_{\bm{\alpha}}\left(\bm x_j,t\right)$, 
$\hat{\bm{u}}_{\bm{\alpha}}^{ij} = \hat{\bm{u}}_{\bm{\alpha}}\left(\bm x_i,t\right) - \hat{\bm{u}}_{\bm{\alpha}}\left(\bm x_j,t\right)$, $D^\alpha W_{ij} = D^\alpha (W)\left(\bm x_i - \bm x_j\right)$, and $V_j = \frac{m_{\bm x_j}}{\rho_{\bm x_j}}$.
We note that
\begin{equation}\label{eq:td_simplify}
    \left(\sum_{\bm x_j \in  N_{\bm x_i}}{ \sum_{\bm{\alpha}\in \mathcal{J}} \hat{\bm{u}}_{\bm{\alpha}}^j \, \Phi_{\bm{\alpha}}(\bm{\xi})}
     W_{ij} V_{j} \right)_t = \sum_{\bm x_j \in  N_{\bm x_i}}{ \sum_{\bm{\alpha}\in \mathcal{J}} \left(\hat{\bm{u}}_{\bm{\alpha}}^j\right)_t \, \Phi_{\bm{\alpha}}(\bm{\xi})}.
     W_{ij} V_{j} ,
\end{equation}
where $\left(\hat{\bm{u}}_{\bm{\alpha}}^j\right)_t$ is the time-derivative of the expansion coefficient $\hat{\bm{u}}_{\bm{\alpha}}^j = \hat{\bm{u}}_{\bm{\alpha}}\left(\bm x_j,t\right)$.
Therefore, Eq. \eqref{eq:spde_sph2} further reduces to
\begin{equation}\label{eq:spde_sph3}
    \sum_{\bm x_j \in  N_{\bm x_i}}{ \sum_{\bm{\alpha}\in \mathcal{J}} \left(\hat{\bm{u}}_{\bm{\alpha}}^j\right)_t \, \Phi_{\bm{\alpha}}(\bm{\xi})}.
     W_{ij} V_{j}  +  \sum_{|\alpha| \leq m} \sum_{\bm x_j \in  N_{\bm x_i}} \sum_{\bm{\alpha}\in \mathcal{J}} \left\{ a_\alpha(\cdot;b) { \hat{\bm{u}}_{\bm{\alpha}}^{ij}\, \Phi_{\bm{\alpha}}(\bm{\xi}) } 
    {D^\alpha W_{ij}}
      V_j\right\}
      = \bm f\left(\cdot,\cdot;\bm \xi\right).
\end{equation}
Finally, we employ Galerkin projection to Eq. \eqref{eq:spde_sph3},
\begin{equation}\label{eq:sph3_Galerkin}
\begin{split}
    & \mathbb E \left[ \left(\sum_{\bm x_j \in  N_{\bm x_i}}{ \sum_{\bm{\alpha}\in \mathcal{J}} \left(\hat{\bm{u}}_{\bm{\alpha}}^j\right)_t \, \Phi_{\bm{\alpha}}(\bm{\xi})}.
     W_{ij} V_{j}  +  \sum_{|\alpha| \leq m} \sum_{\bm x_j \in  N_{\bm x_i}} \sum_{\bm{\alpha}\in \mathcal{J}} \left\{ a_\alpha(\cdot;b) { \hat{\bm{u}}_{\bm{\alpha}}^{ij}\, \Phi_{\bm{\alpha}}(\bm{\xi}) } 
    {D^\alpha W_{ij}}
      V_j\right\} \right) \Phi_{l}(\bm{\xi})\right] \\
      & = \mathbb E \left[\bm f\left(\cdot,\cdot;\bm \xi\right)\Phi_{l}(\bm{\xi})\right], \;\; \forall l=0,\ldots, N,
\end{split}
\end{equation}
where $\mathbb E \left[ \cdot \right]$ is the expectation operator. On further simplifying, we obtain,
\begin{equation}\label{eq:sph3_Galerkin2}
    \begin{split}
        &  \sum_{\bm x_j \in  N_{\bm x_i}}{ \sum_{\bm{\alpha}\in \mathcal{J}} \left(\hat{\bm{u}}_{\bm{\alpha}}^j\right)_t}
     W_{ij} V_{j} \mathbb E \left[\Phi_{\bm{\alpha}}(\bm{\xi})\Phi_{l}(\bm{\xi}) \right] + 
     \sum_{|\alpha| \leq m} \sum_{\bm x_j \in  N_{\bm x_i}} \sum_{\bm{\alpha}\in \mathcal{J}} \left\{ \mathbb E \left[ a_\alpha(\cdot;b) { \hat{\bm{u}}_{\bm{\alpha}}^{ij}\, \Phi_{\bm{\alpha}}(\bm{\xi})\Phi_{l}(\bm{\xi}) }\right] 
    {D^\alpha W_{ij}}
      V_j\right\}  \\
      & = \mathbb E \left[\bm f\left(\cdot,\cdot;\bm \xi\right)\Phi_{l}(\bm{\xi})\right], \;\; \forall l=0,\ldots, N.
    \end{split}
\end{equation}
Assuming the basis functions are orthonormal (a special case of orthogonal basis),
\begin{equation}\label{eq:orthonormality}
    \mathbb E \left[\Phi_{k}(\bm{\xi})\Phi_{l}(\bm{\xi}) \right] = \left\{ \begin{array}{ll}
         1 & \text{if }\,  k=l, \\
         0 & \text{elsewhere.}
    \end{array} \right. 
\end{equation}
Using Eq. \eqref{eq:orthonormality}, Eq. \eqref{eq:sph3_Galerkin2} reduces to,
\begin{equation}\label{eq:sph3_Galerkin3}
    \begin{split}
        &  \sum_{\bm x_j \in  N_{\bm x_i}}{  \left(\hat{\bm{u}}_{l}^j\right)_t}
     W_{ij} V_{j}  + 
     \sum_{|\alpha| \leq m} \sum_{\bm x_j \in  N_{\bm x_i}} \sum_{\bm{\alpha}\in \mathcal{J}} \left\{ \mathbb E \left[ a_\alpha(\cdot;b) { \hat{\bm{u}}_{\bm{\alpha}}^{ij}\, \Phi_{\bm{\alpha}}(\bm{\xi})\Phi_{l}(\bm{\xi}) }\right] 
    {D^\alpha W_{ij}}
      V_j\right\}  \\
      & = \mathbb E \left[\bm f\left(\cdot,\cdot;\bm \xi\right)\Phi_{l}(\bm{\xi})\right], \;\; \forall l=0,\ldots, N.
    \end{split}
\end{equation}
Noting, 
\begin{equation}
    \sum_{\bm x_j \in  N_{\bm x_i}}{  \left(\hat{\bm{u}}_{l}^j\right)_t}
     W_{ij} V_{j} = \left(\hat{\bm{u}}_{l}^i\right)_t,
\end{equation}
where
$\hat{\bm{u}}_{l}^i = \hat{\bm u} \left(\bm x_i,t \right)$, Eq. \eqref{eq:sph3_Galerkin3} reduces to
\begin{equation}\label{eq:sph3_Galerkin4}
    \left(\hat{\bm{u}}_{l}^i\right)_t  + 
     \sum_{|\alpha| \leq m} \sum_{\bm x_j \in  N_{\bm x_i}} \sum_{\bm{\alpha}\in \mathcal{J}} \left\{ \mathbb E \left[ a_\alpha(\cdot;b) { \hat{\bm{u}}_{\bm{\alpha}}^{ij}\, \Phi_{\bm{\alpha}}(\bm{\xi})\Phi_{l}(\bm{\xi}) }\right] 
    {D^\alpha W_{ij}}
      V_j\right\}  = \mathbb E \left[\bm f\left(\cdot,\cdot;\bm \xi\right)\Phi_{l}(\bm{\xi})\right], \;\; \forall l=0,\ldots, N.
\end{equation}
We note that \eqref{eq:sph3_Galerkin4} represents a system of coupled ordinary differential equations in terms of the expansion coefficients. Time integration techniques commonly used in scientific computing can be employed to compute the expansion coefficients. 

\subsection{Boundary and Initial conditions}\label{subsec_bc_ic}
For solving the coupled equation derived in Eq. \eqref{eq:sph3_Galerkin4}, the initial conditions on the expansion coefficients. This is obtained by combining the S-SPH expansion defined in Eq. \eqref{eq:s_sph_ortho} with the initial condition defined in Eq. \eqref{eq:ic}. Substituting Eq. \eqref{eq:s_sph_ortho} into Eq. \eqref{eq:ic}, one obtains,
\begin{equation}\label{eq:ssph_ic1}
    \sum_{\bm x_j \in  N_{\bm x_i}}{ \underbrace{\left\{\sum_{\bm{\alpha}\in \mathcal{J}} \hat{\bm{u}}_{\bm{\alpha}}(\bm{x}_j, t) \, \Phi_{\bm{\alpha}}(\bm{\xi})\right\}}_{\bm u \left( \bm x_j, t; \bm \xi \right)}
     W \left( \bm x_i - \bm x_j; h \right)} \frac{m_{\bm x_j}}{\rho_{\bm x_j}} = \bm u_0\left(\bm x_i; \bm \xi \right).
\end{equation}
Following a similar procedure as Section \ref{subsec:coeff}, we project Eq. \eqref{eq:ssph_ic1} onto the basis function $\Phi_{l}(\bm{\xi})$ to obtain,
\begin{equation}\label{eq:ssph_ic}
    \hat{\bm u}_l\left(\bm x_i \right) = \mathbb E \left[ \bm u_0 \left(\bm x_i; \bm \xi \right)\Phi_{l}(\bm{\xi})\right],\;\; \forall l = 0,1,\ldots, N\;\; \bm x_i \in \mathcal X.
\end{equation}
Given the functional form of $\bm u_0 \left(\cdot; \bm \xi \right)$ is known from Eq. \eqref{eq:ic}, it is trivial to compute the initial condition of the expansion coefficients using Eq. \eqref{eq:ssph_ic}. Similarly, the boundary conditions of the expansion coefficients are given as
\begin{equation}\label{eq:ssph_bcd}
    \hat{\bm{u}}_l(\bm{x}_i,t) = \mathbb{E}[\bm{g_1}(\bm{x}_i,t,\bm{\xi})\Phi_l(\bm{\xi})], \quad \text{on } \partial \mathcal{X}_D \times [0, T], \quad \forall l=0,1,\ldots,N.
\end{equation}
A similar expression can be derived for the Neumann boundary condition as well. However, in practice, the Neumann boundary condition is trivially satisfied during Galerkin projection and hence, the same is not shown here.

Given the kernel-based expression of S-SPH in Eq. \eqref{eq:s_sph_ortho}, the challenges associated with the treatment of the boundary condition and handling of free surfaces for conventional SPH are also present for the proposed framework. To address these, we employ well-known techniques such as gradient correction \cite{chen1999corrective} in S-SPH as well.

\subsection{Gradient correction}
In S-SPH, handling boundary conditions accurately is often challenging due to issues like particle clustering or insufficient particle support near boundaries. One effective approach to improve the accuracy of SPH simulations near boundaries is gradient correction, which adjusts the gradient operator to better represent the approximations by improving their consistency. This helps in achieving $C^0$ consistency near the boundary. Moreover, this also improves the consistency in the interior domain ($C^1$). This method enhances the accuracy of derivative calculations, particularly in regions with irregular and insufficient particle distributions. The gradient correction modifies the gradient operator by introducing a correction matrix \( \mathbf{B}_j \), which adjusts the kernel gradient based on the local particle configuration. The corrected gradient for an expansion coefficient \( \hat {u}_l\) at particle position \( \bm x_j \) is expressed as,
\begin{equation}\label{eq:gradient_correction}
    \nabla \hat u_l^j = \mathbf{B}_j \sum_k m_k \frac{{\hat u}_l^k - {\hat u}_l^j}{\rho_k} \nabla W(\bm{x}_j - \bm{x}_k; h),
\end{equation}
where $\hat u_l^j = \hat u_l (\bm x_j,t). $ \( \mathbf{B}_j \) in Eq. \eqref{eq:gradient_correction} is the gradient correction matrix for $j-$th particle, and is calculated as,
\begin{equation}
    \mathbf{B}_j = \left( \sum_k m_k \frac{(\bm{x}_k - \bm{x}_j) \otimes \nabla W(\bm{x}_j - \bm{x}_k, h)}{\rho_k} \right)^{-1}.
\end{equation}
In essence, the matrix \( \mathbf{B}_j \) effectively rescales the gradient based on the distribution of particles around \( j \), helping to account for the asymmetry in support near boundaries.

\subsection{Karhunen Lo{\`e}ve expansion}
As previously noted, the uncertainty in the SPDE presented in Eq. \eqref{eq:spde} may originate from several sources, including the system parameter $b(\bm{\xi})$, the forcing function $f\left(\bm{x}, t; \bm{\xi} \right)$, the boundary conditions $g_1 \left(\bm{x}, t; \bm{\xi} \right)$ and $g_2 \left(\bm{x}, t; \bm{\xi} \right)$, as well as the initial condition $\bm{u}_0 \left(\bm{x}; \bm{\xi} \right)$. These quantities, in addition to being stochastic, are frequently dependent on spatial and/or temporal variables and are thus appropriately modeled as random fields or stochastic processes. Under such circumstances, it becomes imperative to reformulate the associated random fields in terms of a finite set of random variables. Established methodologies for this purpose include point discretization \cite{ghanem2003stochastic,xiu2010numerical}, the average discretization method~\cite{shinozuka1991simulation,grigoriu2013stochastic}, and series expansion techniques~\cite{todor2007convergence,babuska2004galerkin}.
While the proposed approach is compatible with any of these discretization strategies, the Karhunen–Lo{\`e}ve Expansion (KLE) is employed herein owing to its desirable convergence characteristics.

Consider $\kappa \left(\bm x;\bm \xi\right)$ to be a generic random field. For simplicity, we have assumed the stochastic field $\kappa$ to only vary with space. As previously stated, $\kappa$ can potentially represent the system parameter, initial and boundary conditions, and the forcing function. Using KLE, the random field $\kappa \left(\bm x; \bm \xi\right)$ can be represented as,
\begin{equation}\label{eq:kle}
    \kappa(\bm{x},\bm{\xi}) = \bar{\kappa}(\bm{x}) + \sum_{k=1}^{M} \sqrt{\lambda_k} \psi_k(\bm{x}) \xi_k,
\end{equation}
where \( \bar{\kappa}(\bm{x}) = \mathbb{E}[\kappa(\bm{x};\bm{\xi})] \) denotes the mean function and \( \lambda_k \), \( \phi_k(\bm{x}) \) are the eigenvalues and eigenfunctions of the covariance kernel \( C_{\kappa}(\bm{x},\bm{x}') \), satisfying the integral eigenvalue problem,
\begin{equation}\label{eq:eigen}
\int_{\mathcal{X}} C_{\kappa}(\bm{x},\bm{x}') \psi_k(\bm{x}') \, d\bm{x}' = \lambda_k \psi_k(\bm{x}).
\end{equation}
Here, \( \xi_k \) are independent standard normal random variables, and the truncation index \( M \) determines the approximation accuracy. 
Note that the Eigenvalue problem in Eq. \eqref{eq:eigen} is often not solvable exactly, and numerical techniques needs to be employed to solve the same. Nonetheless, the expansion in Eq. \eqref{eq:kle} allows us the discretize the random field into random variable in a seamless manner.

\subsection{Algorithm}
We conclude this section by discussing the algorithm of the proposed S-SPH. This includes generating neighborhood list, computing gradients using the SPH kernel, reducing the underlying SPDE into coupled PDEs by using Galerkin projection, and solving the same using suitable time-integration scheme.
A list of neighboring points is generated based on a user-defined influence radius $c$. This is depicted in 
Algorithm \ref{algo:generate_neighbors}.
\begin{algorithm}[htbp]
  \caption{GenerateNeighborList}
  \label{algo:generate_neighbors}
  \textbf{Input:} 
    Particle positions $\{x_j\}_{j=1}^J$, search radius $c$.\\
  \textbf{Return:} 
    Neighbor lists $\mathcal{N}$, where $\mathcal{N}[j]$ is the set of all $k$ such that $|x_j - x_k| \le c$.
  \begin{algorithmic}[1]
    \State Initialize $\mathcal{N}[1 \dots J] \gets \emptyset$
    \For{$j = 1,\dots,J$}
      \For{$k = 1,\dots,J$}
        \If{$|x_j - x_k| \le c$}
          \State Add $k$ to $\mathcal{N}[j]$
        \EndIf
      \EndFor
    \EndFor
  \end{algorithmic}
\end{algorithm}
The spatial derivatives are calculated by differentiating the S-SPH kernel using automatic differentiation (Algorithm \ref{algo:sph_derivative}). Higher order derivatives can also be obtained by repeatedly using Algorithm \ref{algo:sph_derivative}.
\begin{algorithm}[htbp]
  \caption{SPHDerivative}
  \label{algo:sph_derivative}
  \textbf{Input:} 
    Scalar field values $\{u_{l k}\}_{k=1}^J$ at polynomial level $l$, masses $\{m_k\}$, densities $\{\rho_k\}$,  
    target index $j$, neighbor list $\mathcal{N}_j$, kernel $W(\cdot,h)$.\\
  \textbf{Return:} 
    SPH derivative $D$ at particle $j$.
  \begin{algorithmic}[1]
    \State $D \gets 0$
    \For{each $k \in \mathcal{N}_j$}
      \State Compute 
        $\displaystyle \nabla W_{jk} \;=\; \frac{\partial W(x_j - x_k,\,h)}{\partial x_k}$
      \State Update
        $\displaystyle D \;\gets\; D \;+\; m_k \,\frac{u_{l k} - u_{l j}}{\rho_k}\;\nabla W_{jk}$
    \EndFor
    \State \Return $D$
  \end{algorithmic}
\end{algorithm}
Similar to conventional SPH, we use the predictor-corrector scheme for time marching. This allows better stability and superior results. The overall S-SPH algorithm using the S-SPH is shown in Algorithm \ref{alg:galerkin-sph-main}, which exploits  Algorithms \ref{algo:generate_neighbors} and \ref{algo:sph_derivative} discussed before for computing the neighborhood list and the spatial gradients as discussed before. 
\begin{algorithm}
\caption{Stochastic‑SPH with Galerkin‑Coupled Predictor–Corrector Scheme}
\label{alg:galerkin-sph-main}
\begin{algorithmic}[1]
\Require Number of particles $J$, positions $x_j$, masses $m_j$, densities $\rho_j$, radius $c$, smoothing $h$, time step $\Delta t$, steps $N$, PCE size $P$, max order $m$
\Ensure Expansion coefficients $\hat u_{j\ell}^n$ and positions $x_{j}^n$
\State \textbf{Precompute}
\State \quad Call Algorithm~\ref{alg:compute-coupling-coeffs} to obtain $C_{\alpha\ell}$
\State \quad Compute forcing projections $F_j^\ell = \mathbb{E}[f(x_j,t;\xi)\,\Phi_\ell(\xi)]$
\State \quad Initialize $\hat u_{j\ell}^0 = \mathbb{E}[u_0(x_j;\xi)\,\Phi_\ell(\xi)]$, $x_j^0 = x_j$
\For{$n=0,\dots,N-1$}  \Comment{Time‐integration loop}
  \State $N_j = \mathrm{GenerateNeighborList}(\{x_k^n\},c)$ \; $\forall j$ \Comment{Algorithm \ref{algo:generate_neighbors}}
  \For{$j=1,\dots,J$}  \Comment{Predictor step}
    \For{$\ell=0,\dots,P$}
      \State $d_{j\alpha}^n \gets \mathrm{SPHDerivative}\bigl(\{\hat u_{k\alpha}^n\}_{k=1}^J,\;\alpha,\;j,\;N_j\bigr)\quad\forall\,\alpha:|\alpha|\le m$ \Comment{Algorithm \ref{algo:sph_derivative}}
      \State $R_{j\ell}^n \gets -\sum_{|\alpha|\le m}C_{\alpha\ell}\,d_{j\alpha}^n + F_j^\ell$
      \State $\tilde u_{j\ell}^{\,n+1} \gets \hat u_{j\ell}^n + \Delta t\,R_{j\ell}^n$
    \EndFor
    \State $\tilde x_j^{\,n+1} \gets x_j^n + \Delta t\,\hat u_{j0}^n$
  \EndFor
  \State $\tilde N_j = \mathrm{GenerateNeighborList}(\{\tilde x_k^{n+1}\},c)$ \; $\forall j$ \Comment{Algorithm \ref{algo:generate_neighbors}}
  \For{$j=1,\dots,J$}  \Comment{Corrector step}
    \For{$\ell=0,\dots,P$}
      \State $\tilde d_{j\alpha}^{\,n+1} \gets \mathrm{SPHDerivative}\bigl(\{\tilde u_{k\alpha}^{\,n+1}\},\;\alpha,\;j,\;\tilde N_j\bigr)\quad\forall\,\alpha:|\alpha|\le m$ \Comment{Algorithm \ref{algo:sph_derivative}}
      \State $\tilde R_{j\ell}^{\,n+1} \gets -\sum_{|\alpha|\le m}C_{\alpha\ell}\,\tilde d_{j\alpha}^{\,n+1} + F_j^\ell$
      \State $\hat u_{j\ell}^{\,n+1} \gets \hat u_{j\ell}^n + \tfrac{\Delta t}{2}\bigl(R_{j\ell}^n + \tilde R_{j\ell}^{\,n+1}\bigr)$
    \EndFor
    \State $x_j^{\,n+1} \gets x_j^n + \tfrac{\Delta t}{2}\bigl(\hat u_{j0}^n + \hat u_{j0}^{\,n+1}\bigr)$
  \EndFor
\EndFor
\end{algorithmic}
\end{algorithm}
The coupling coefficients for reducing the underlying S-PDE into coupled PDEs can either be computed in closed form by hand calculation or by using Algorithm \ref{alg:compute-coupling-coeffs}. The expectation in Algorithm \ref{alg:compute-coupling-coeffs} is computed using the orthogonal property and numerical integration.
\begin{algorithm}
\caption{Compute Galerkin Coupling Coefficients}
\label{alg:compute-coupling-coeffs}
\begin{algorithmic}[1]
\Require Multi‐index set $\mathcal{J}=\{\alpha:|\alpha|\le q\}$, orthonormal basis $\{\Phi_\alpha\}$, operator coeffs.\ $\{a_\alpha(\xi)\}$, max coupling order $m$, max basis index $N$
\Ensure Coupling coefficients $C_{\alpha\ell}$ for all $|\alpha|\le m$, $\ell=0,\dots,N$
\ForAll{$\alpha\in\mathcal{J}$ with $|\alpha|\le m$}
  \For{$\ell=0,\ldots,N$}
    \State $C_{\alpha\ell}\;\gets\;\mathbb{E}\bigl[a_\alpha(\xi)\,\Phi_\alpha(\xi)\,\Phi_\ell(\xi)\bigr]$ \Comment{Numerical integration/Orthogonal property}
  \EndFor
\EndFor
\end{algorithmic}
\end{algorithm}

\section{Numerical examples}\label{sec:numerical}
In this section, we present three representative benchmark problems to demonstrate the effectiveness and robustness of the proposed approach. These examples encompass systems governed by both linear and nonlinear partial differential equations (PDEs), each incorporating various sources of uncertainty, including randomness in system parameters and initial conditions. The selected benchmarks are designed to capture diverse physical scenarios and mathematical complexities, thereby providing a comprehensive evaluation of our method.
We have conducted detailed case studies focusing on two key aspects: (1) the impact of the basis function order on convergence behavior, and (2) the influence of different types of randomness -- such as parametric uncertainty and stochastic initial fields. These studies are intended to assess the method’s adaptability, accuracy, and scalability across a range of uncertainty configurations.
To quantify performance, the results obtained using the proposed approach are systematically compared against reference solutions generated via vanilla Monte Carlo Simulation (MCS), which serves as the baseline. Both accuracy and computational efficiency are evaluated, highlighting the advantages of our method in terms of faster convergence, and lower computational cost while maintaining high fidelity in the solution statistics.

\subsection{Example 1: Stochastic Wave Advection Equation}
As the first example, we consider the well know wave advection equation. This is a first-order hyperbolic partial differential equation that models the movement of scalar field subjected to a known velocity field. The governing equation for the wave advection equation takes the following form:
\begin{equation}
    \frac{\partial u}{\partial t} = c \frac{\partial u}{\partial x} \quad \quad c\in \mathbb{R}, \text{for x} \in (0,1), \text{and t} > 0 
\end{equation}
A periodic boundary condition is considered. 
For the first set of studies, we consider the parameter $c$ to be stochastic and model it as a random variable. We consider two cases: (a) $c \sim \mathcal N \left( 0.06,0.1\right)$ follows a Gaussian distribution, 
and (b) $c \sim \mathcal {LN} \left( 0.06,0.1\right)$ follows lognormal distribution. 
\begin{figure}[ht!]
  \centering

  \begin{tabular}{@{}c@{\hspace{1em}}c@{}}
    \subcaptionbox{Mean S-SPH\label{fig:mean_ssph_2}}{%
      \includegraphics[width=0.48\textwidth]{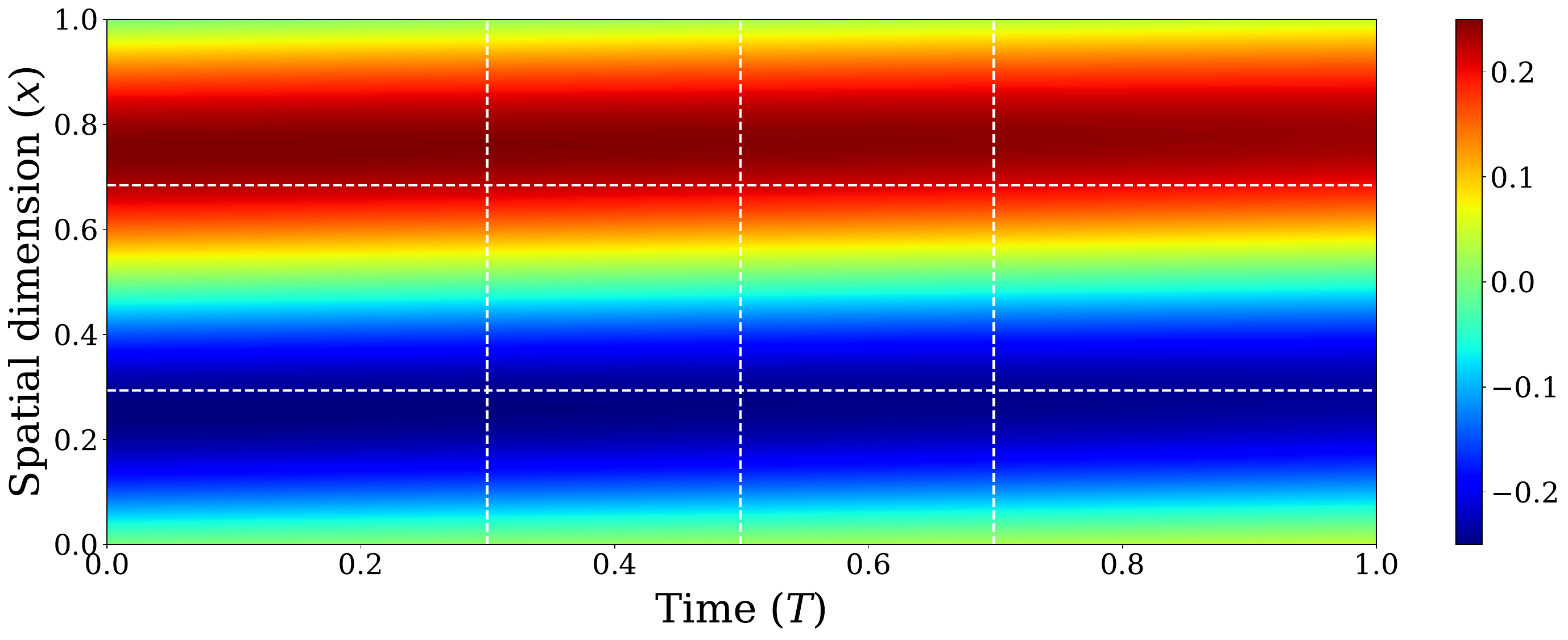}} &
    \subcaptionbox{Mean MCS\label{fig:mean_mcs}}{%
      \includegraphics[width=0.48\textwidth]{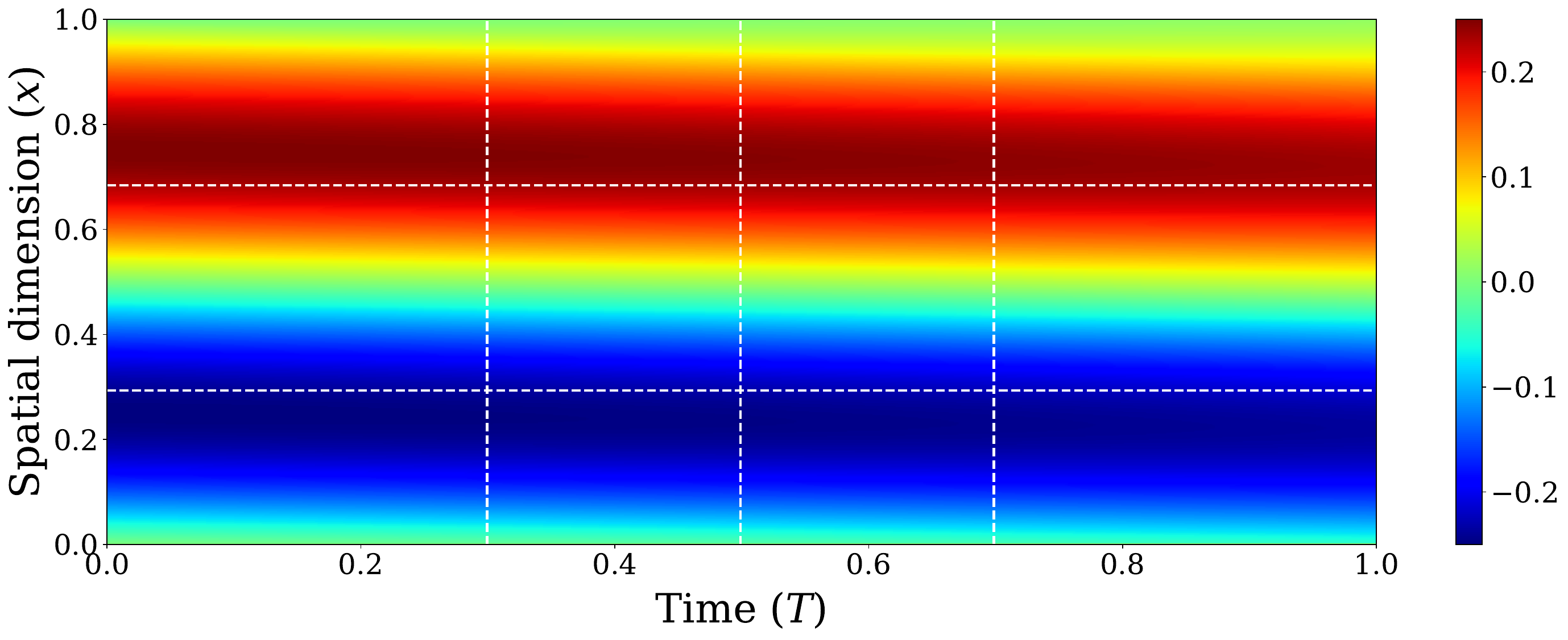}}
  \end{tabular}
  \vspace{1em}
  \begin{tabular}{@{}c@{\hspace{1em}}c@{}}
    \subcaptionbox{Std S-SPH\label{fig:std_ssph}}{%
      \includegraphics[width=0.48\textwidth]{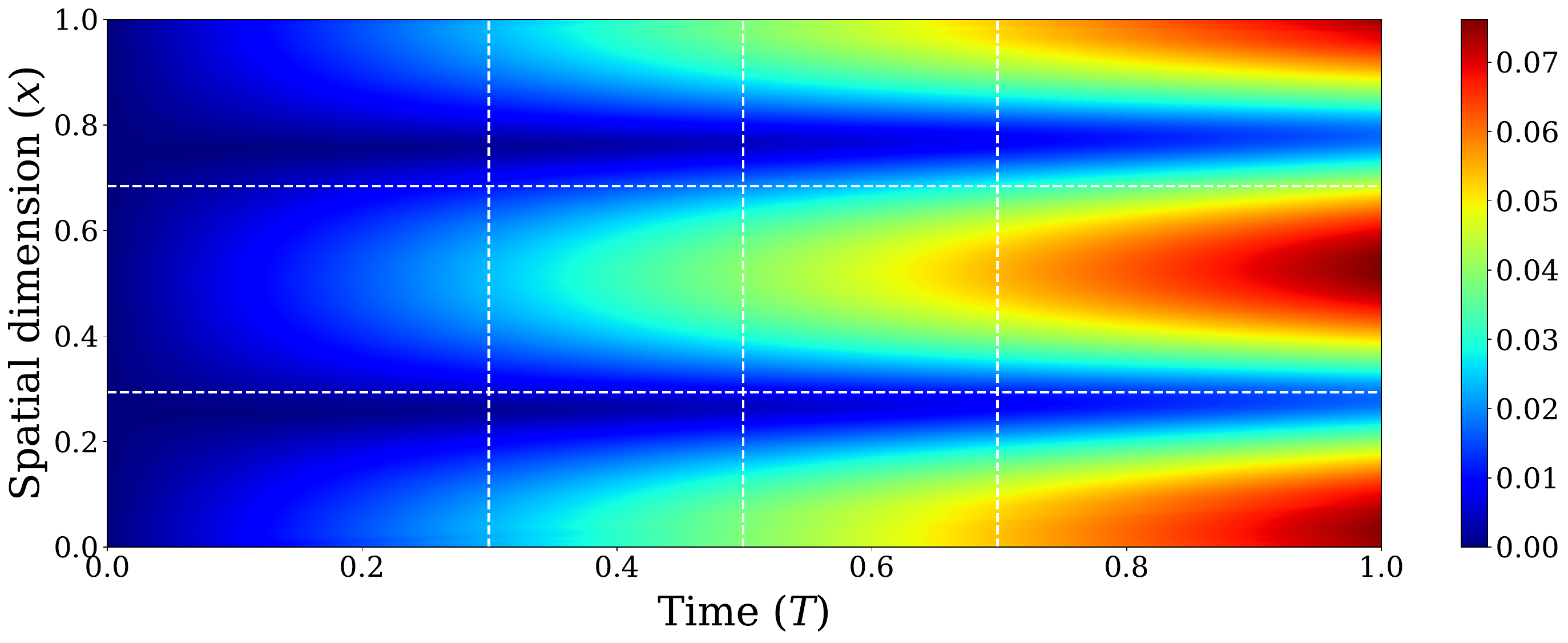}} &
    \subcaptionbox{Std MCS\label{fig:std_mcs}}{%
      \includegraphics[width=0.48\textwidth]{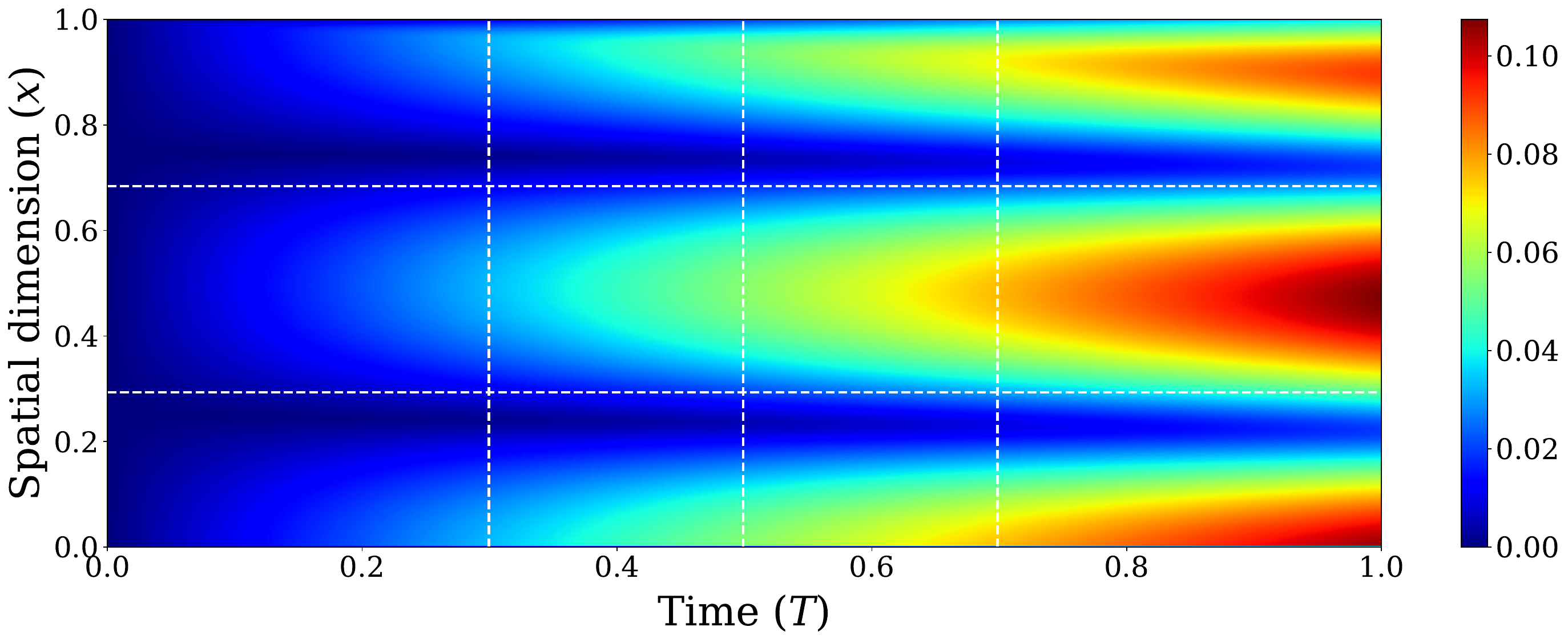}}
  \end{tabular}
  \caption{Mean and standard deviation contours of the solution field for the stochastic wave advection equation with random advection speed $c \sim \mathcal{N}(0.06, 0.1)$. 
  Subfigures (a) and (b) show the spatio–temporal evolution of the mean response obtained using the proposed S-SPH method and MCS, respectively. 
  Subfigures (c) and (d) present the corresponding standard deviation.}
  \label{fig:mean_std_normal}
\end{figure}
\begin{figure}[t]
  \centering

  \begin{tabular}{@{}c@{\hspace{1em}}c@{}}
    \subcaptionbox{Mean S-SPH\label{fig:mean_ssph_1}}{%
      \includegraphics[width=0.48\textwidth]{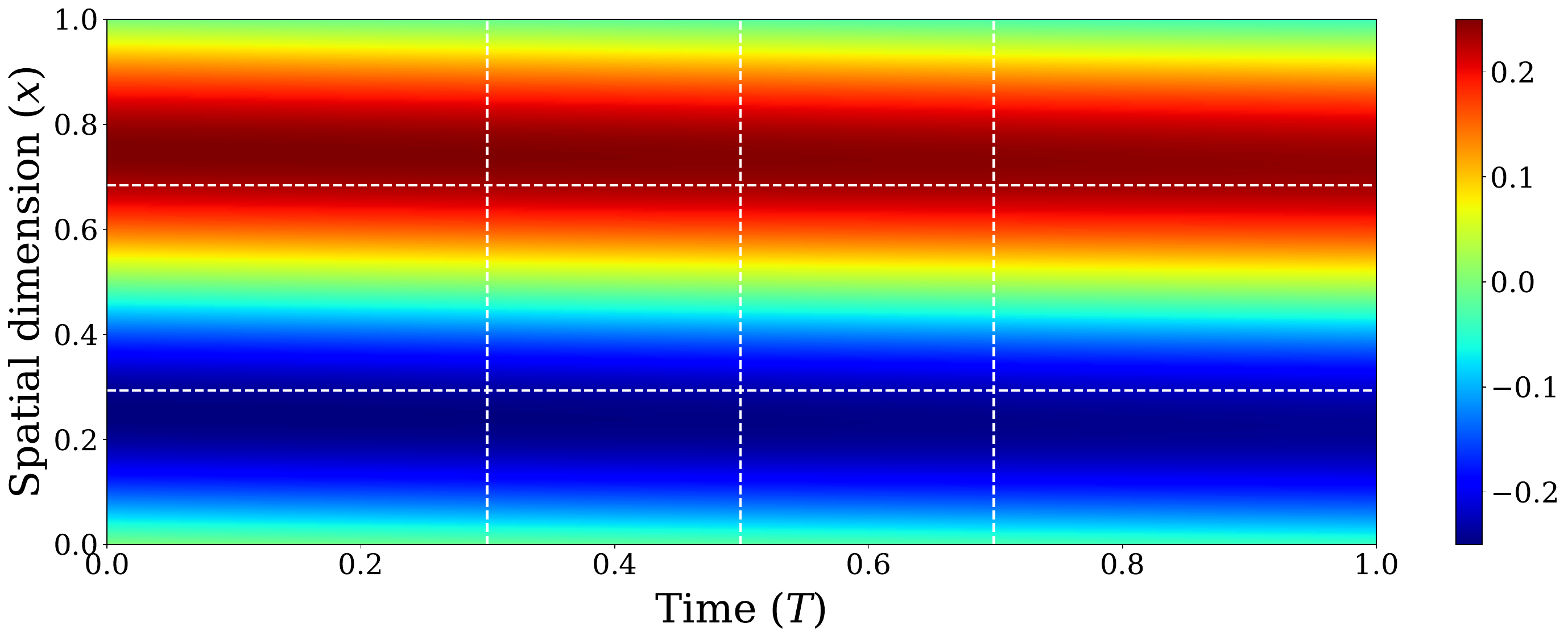}} &
    \subcaptionbox{Mean MCS\label{fig:mean_mcs}}{%
      \includegraphics[width=0.48\textwidth]{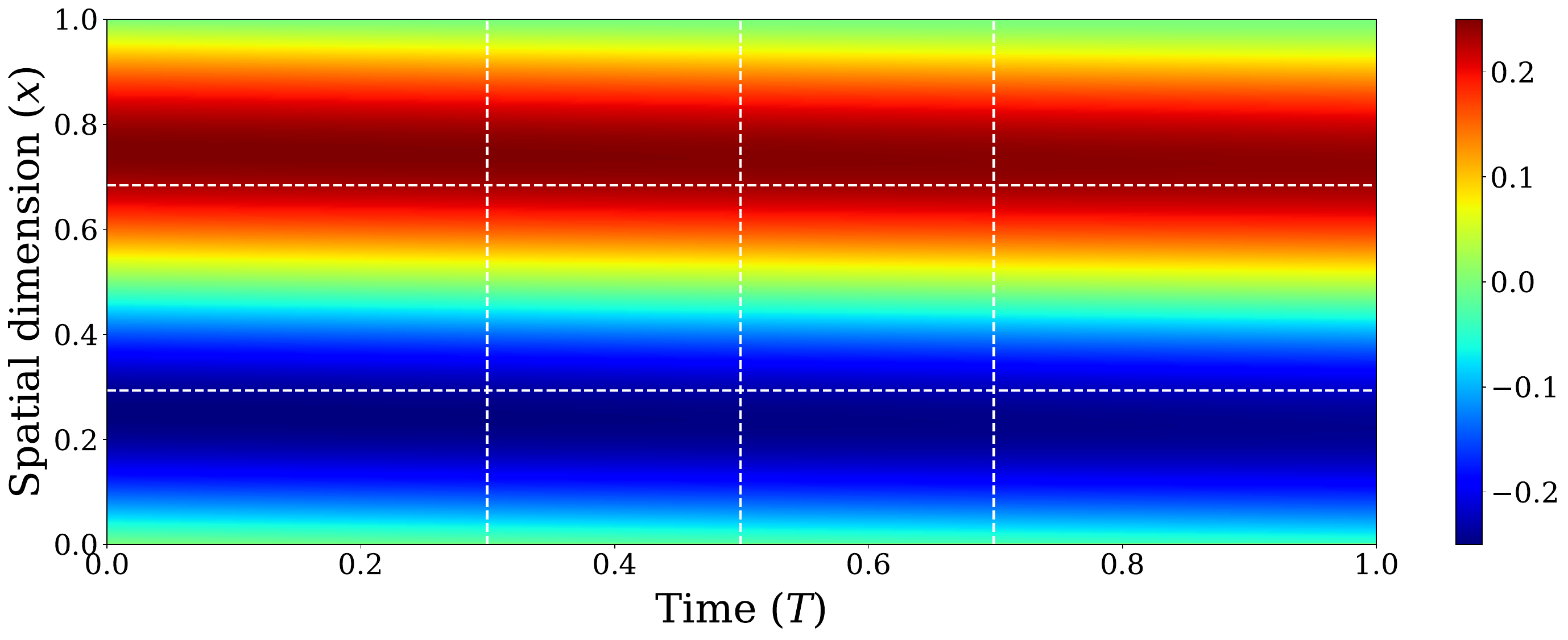}}
  \end{tabular}

  \vspace{1em}
  \begin{tabular}{@{}c@{\hspace{1em}}c@{}}
    \subcaptionbox{Std S-SPH\label{fig:std_ssph}}{%
      \includegraphics[width=0.48\textwidth]{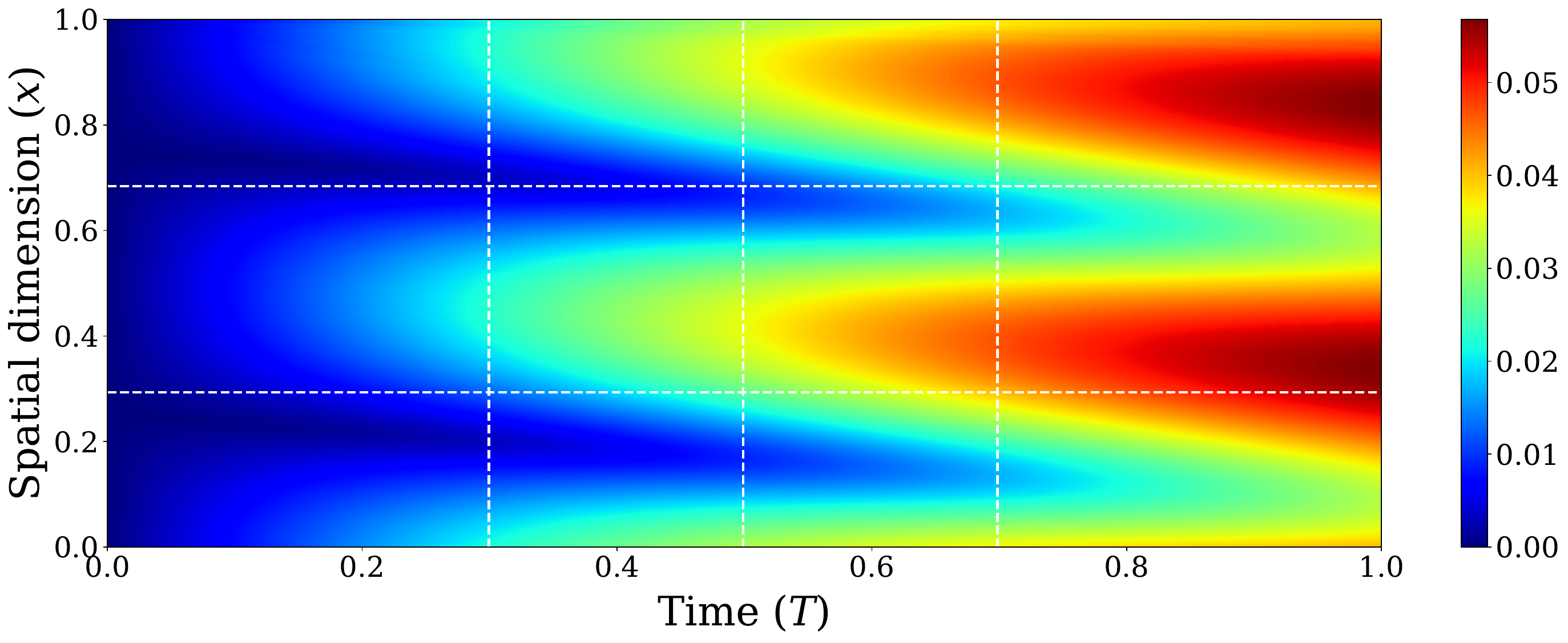}} &
    \subcaptionbox{Std MCS\label{fig:std_mcs}}{%
      \includegraphics[width=0.48\textwidth]{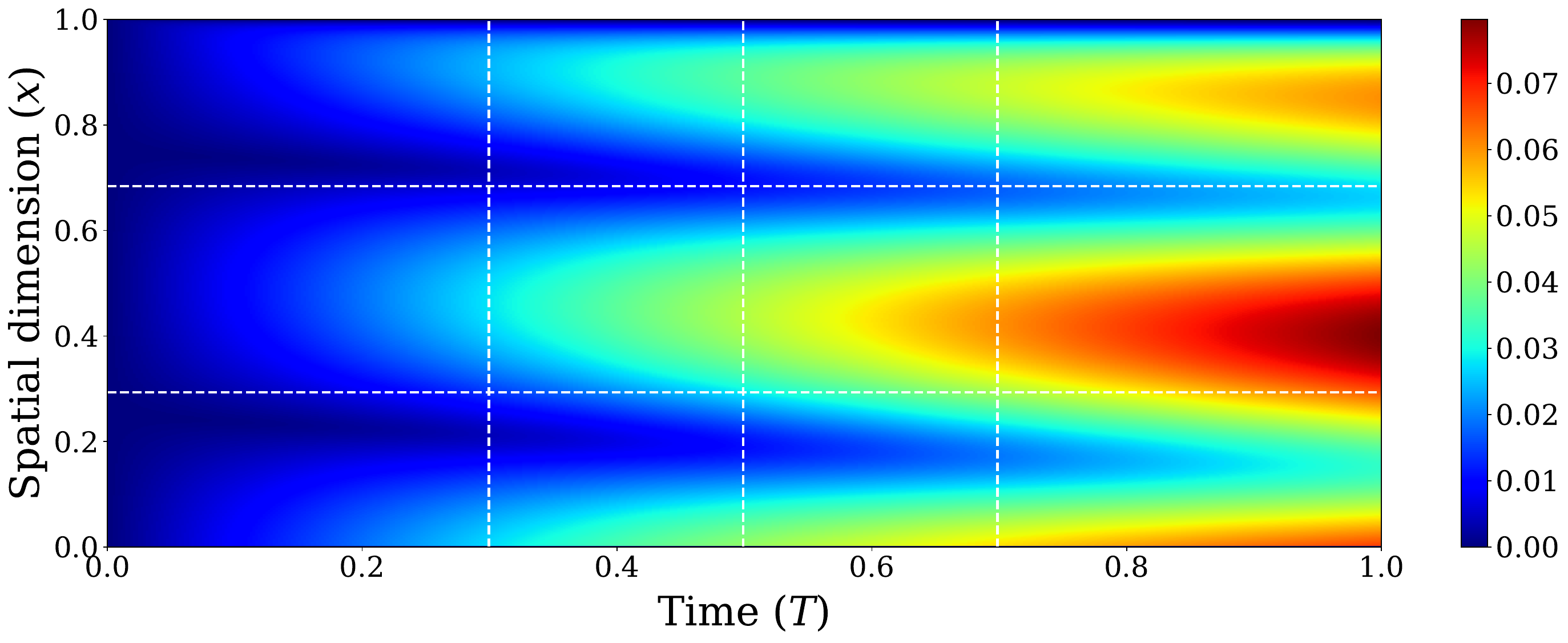}}
  \end{tabular}
  \caption{Spatio-temporal contours of the mean and standard deviation of the solution field for the stochastic wave advection problem with a lognormally distributed advection speed, $c \sim \mathcal{LN}(0.06, 0.1)$. 
 The top row compares the mean response obtained using the S-SPH formulation and MCS, while the bottom row shows the corresponding standard deviation fields.}
  \label{fig:mean_std_lognormal}
\end{figure}
The objective here is to investigate the performance of the proposed S-SPH in capturing the statistics of the response variable. To that end, we employ the proposed S-SPH approach with maximum basis function order $P=5$, time step $\Delta t=0.001$, total integration time $T=0.5$, spatial domain length $a=1$, number of grid points $J=512$, grid spacing $dx=1/J$, smoothing length $h=1.2dx$, influence radius $r=2h$, and $N=T/\Delta t$ time steps. To generate the benchmark solution, MCS has been employed with 5,000 samples drawn from the corresponding distribution.

Figs.~\ref{fig:mean_std_normal} and~\ref{fig:mean_std_lognormal} present the spatio-temporal contours of the mean and standard deviation of the response for the two stochastic cases, as obtained using the proposed S-SPH framework and Monte Carlo simulation (MCS). For both uncertainty models, the mean response predicted by S-SPH shows an almost exact agreement with the MCS reference across the entire space–time domain, indicating negligible bias in the estimated first-order statistics. Furthermore, the standard deviation fields computed using S-SPH closely match those obtained from MCS, demonstrating that the proposed method accurately captures the propagation of uncertainty and reliably reproduces second-order statistical characteristics of the solution.
Overall, these plots are indicative of the capability offered by S-SPH in capturing the stochastic response of stochastic partial differential equations.

In the second set of numerical studies, the parameter \( c \) is modeled as a Gaussian random variable, \( c \sim \mathcal{N}(0.06,\,0.1) \), and the uncertainty is further introduced through the initial condition. Two distinct stochastic representations of the initial condition are considered.
In the first case, the initial condition is parameterized as
\begin{equation}
u_0(x) = \alpha \sin(\beta x),
\end{equation}
where the amplitude \( \alpha \sim \mathcal{N}(0.25,\,0.1) \) and the wavenumber \( \beta \sim \mathcal{N}(2\pi,\,0.1) \) are modeled as independent Gaussian random variables.
In the second, more general case, the initial condition is modeled as a Gaussian random field with a mean function
\begin{equation}
\mu_{u_0}(x) = 0.01 \sin(2\pi x),
\end{equation}
and covariance kernel
\begin{equation}\label{eq:covariance_kernel_rf}
k_{u_0}(x,x') = \sigma_{u_0}^2 
\exp\!\left(-\frac{\|x - x'\|_2^2}{l_{u_0}^2}\right),
\end{equation}
where \( \sigma_{u_0}^2 \) denotes the process variance and \( l_{u_0} \) is the characteristic length scale. For this study, the parameters are set to \( \sigma_{u_0}^2 = 0.001 \) and \( l_{u_0} = 0.01 \).
To obtain a finite-dimensional representation of the random field, the Karhunen--Lo\`eve (KL) expansion described in Eq.~\eqref{eq:kle} is employed. The expansion is truncated by retaining the first five KL modes, which together account for approximately \(95\%\) of the total energy of the stochastic process.

\begin{figure}[t]
  \centering

  \begin{tabular}{@{}c@{\hspace{1em}}c@{}}
    \subcaptionbox{Mean S-SPH\label{fig:mean_ssph_case4}}{%
      \includegraphics[width=0.45\textwidth]{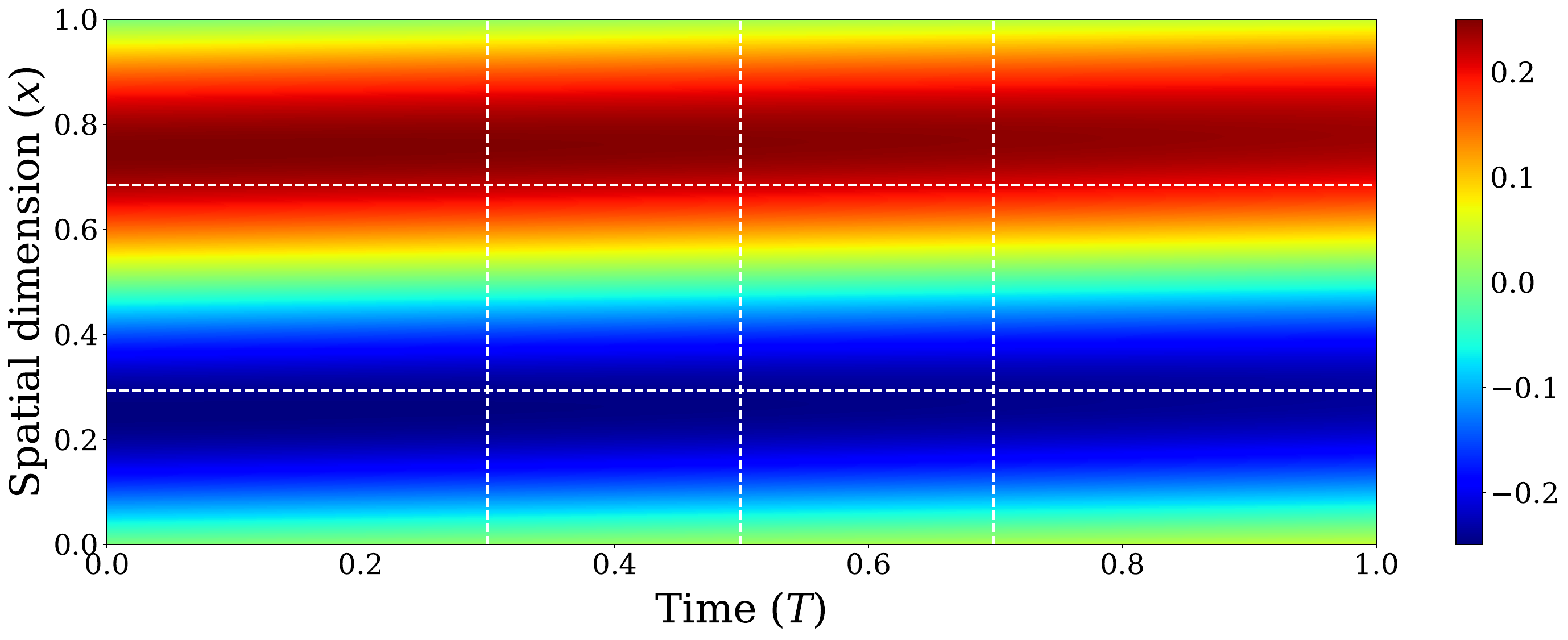}} &
    \subcaptionbox{Mean MCS\label{fig:mean_mcs_case4}}{%
      \includegraphics[width=0.45\textwidth]{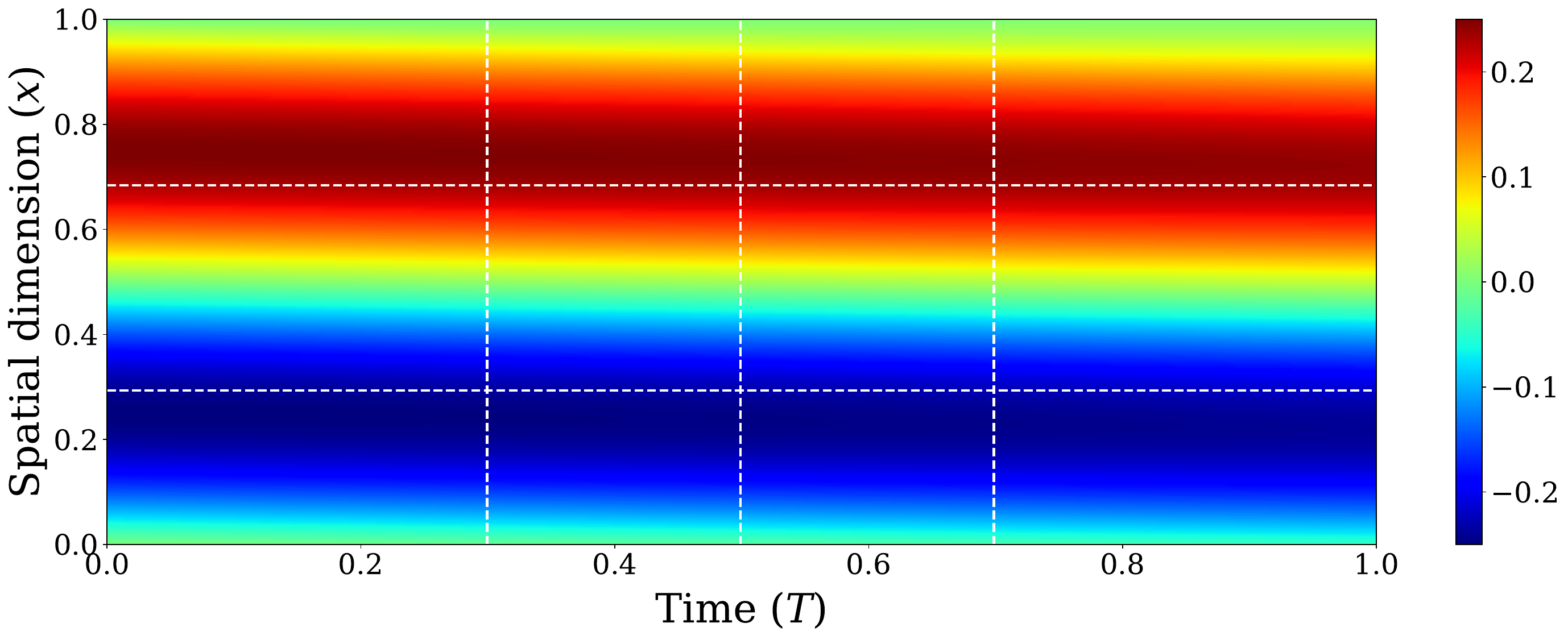}}
  \end{tabular}

  \vspace{1em}

  \begin{tabular}{@{}c@{\hspace{1em}}c@{}}
    \subcaptionbox{Std S-SPH\label{fig:std_ssph_case4}}{%
      \includegraphics[width=0.45\textwidth]{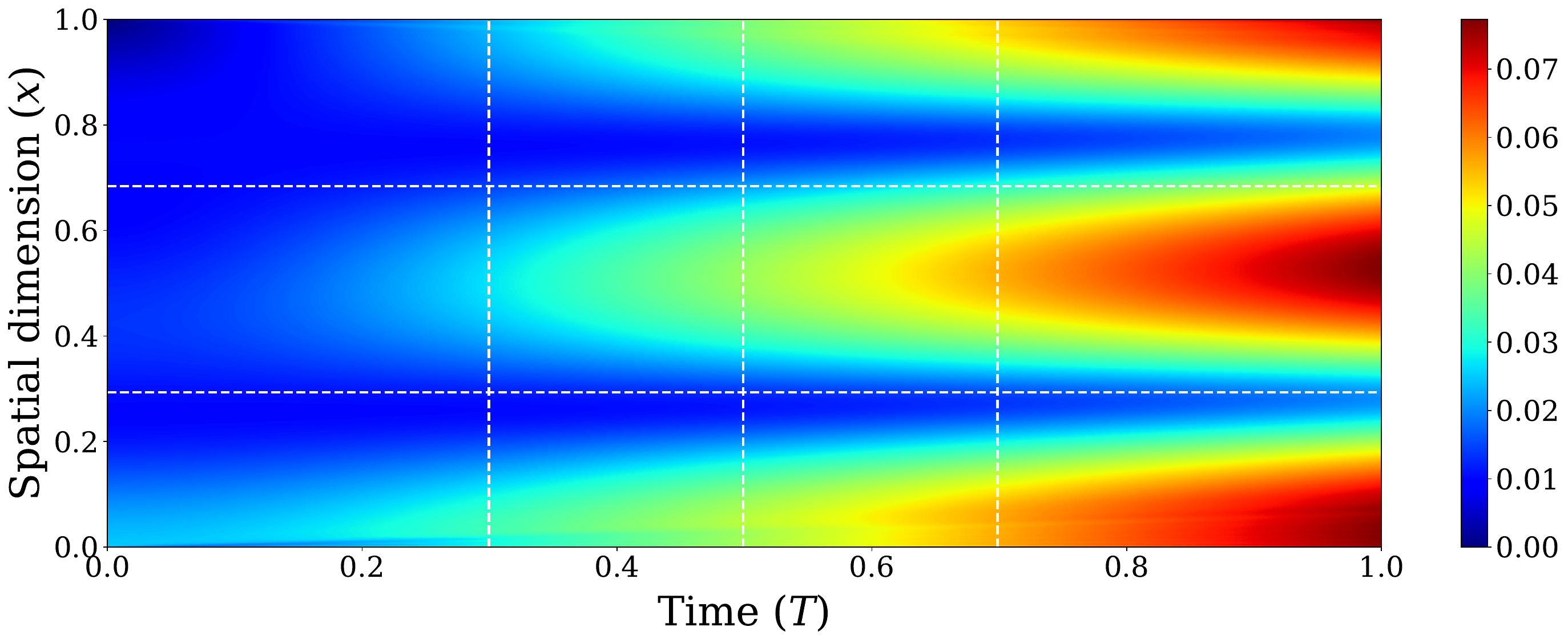}} &
    \subcaptionbox{Std MCS\label{fig:std_mcs_case4}}{%
      \includegraphics[width=0.45\textwidth]{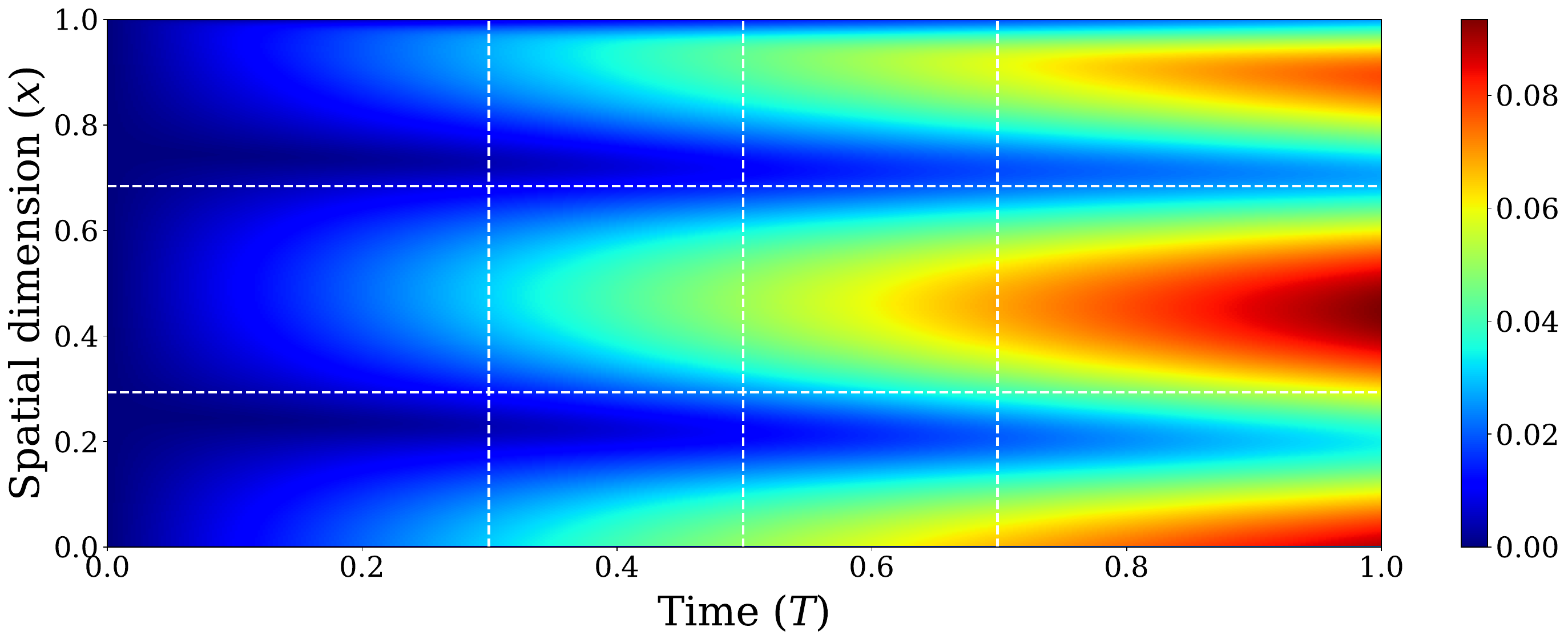}}
  \end{tabular}

  \caption{Mean and standard deviation fields of the solution to the stochastic wave advection equation corresponding to a stochastic initial condition $u_0 = \alpha \sin(\beta x)$, with $\alpha$ and $\beta$ treated as random variables. 
  The mean response predicted by the S-SPH formulation is compared against Monte Carlo Simulation (MCS) in the top row, while the bottom row presents the associated standard deviation profiles.}
  \label{fig:mean_std_u0_sin}
\end{figure}

\begin{figure}[ht!]
  \centering

  \begin{tabular}{@{}c@{\hspace{1em}}c@{}}
    \includegraphics[width=0.65\textwidth]{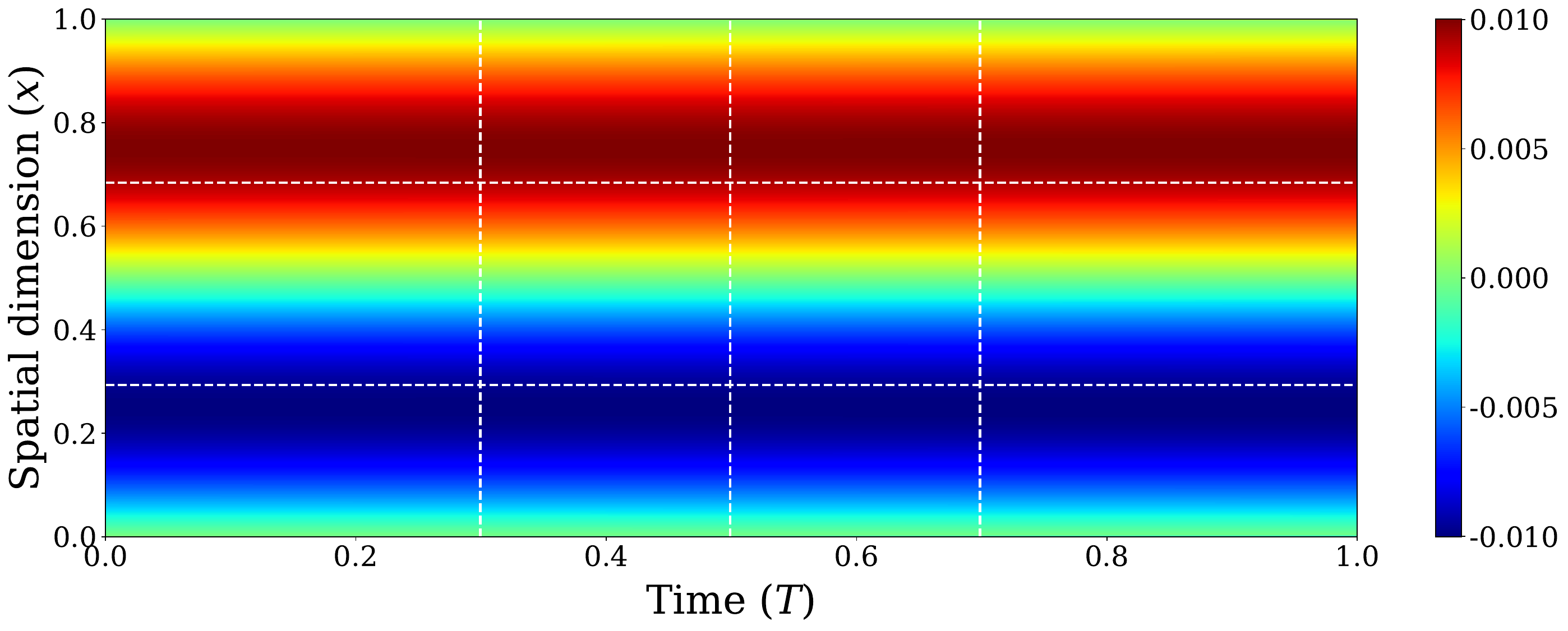} &
    \includegraphics[width=0.30\textwidth]{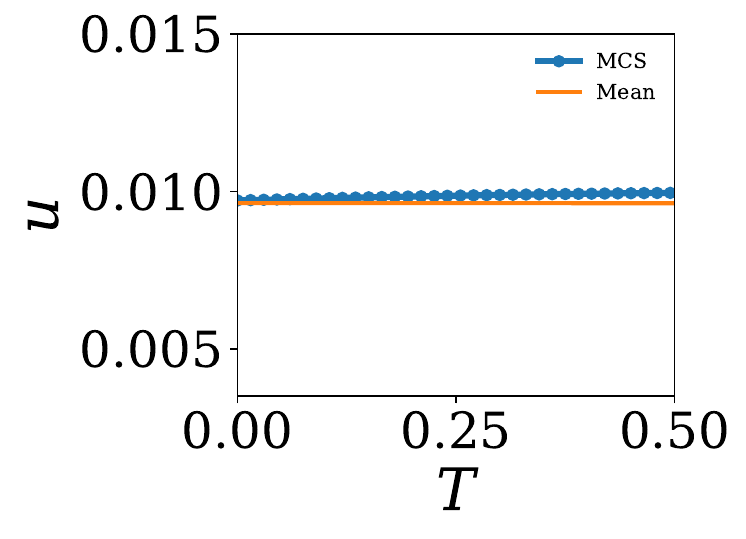}
  \end{tabular}

  \vspace{1em}

  \begin{tabular}{@{}c@{\hspace{1em}}c@{}}
    \includegraphics[width=0.65\textwidth]{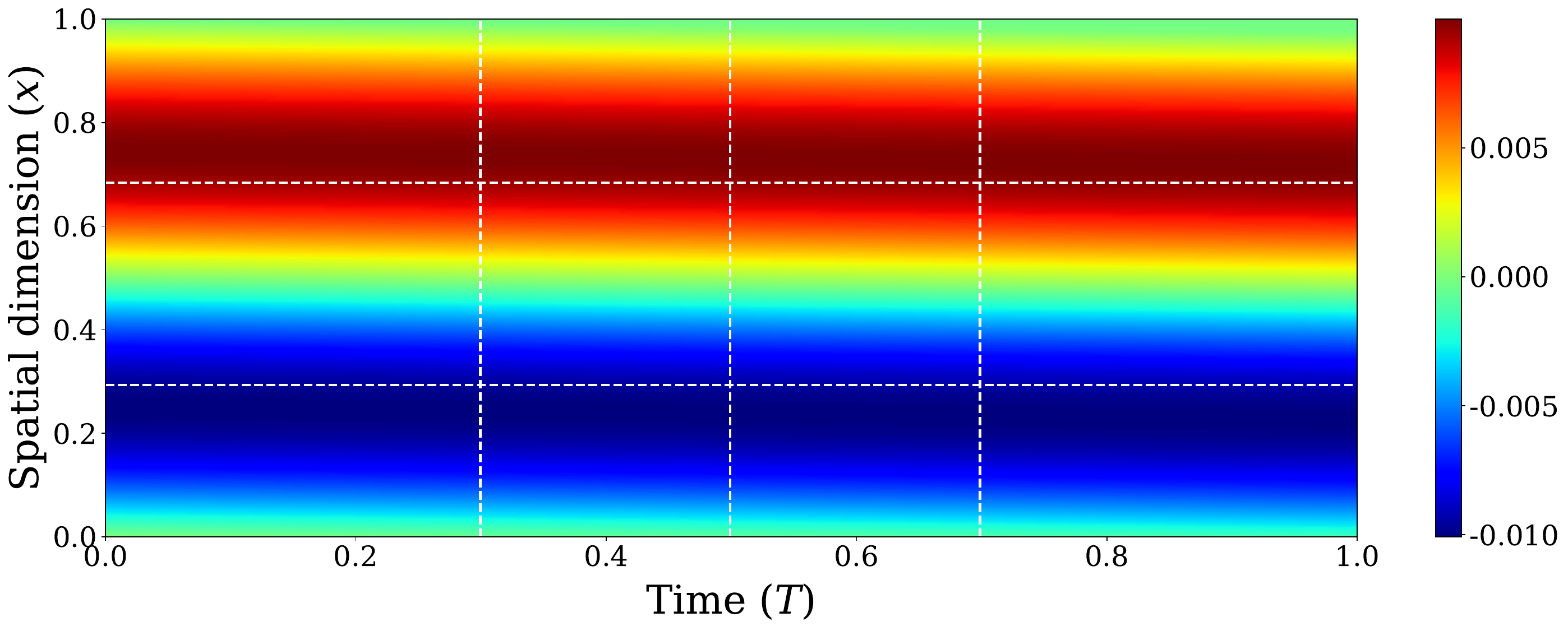} &
    \includegraphics[width=0.30\textwidth]{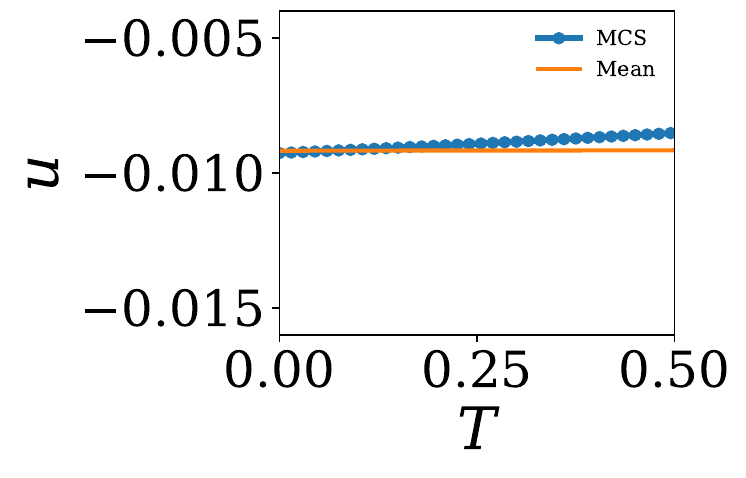}
  \end{tabular}

  \vspace{1em}

  \begin{tabular}{@{}c@{\hspace{1em}}c@{\hspace{1em}}c@{}}
    \includegraphics[width=0.30\textwidth]{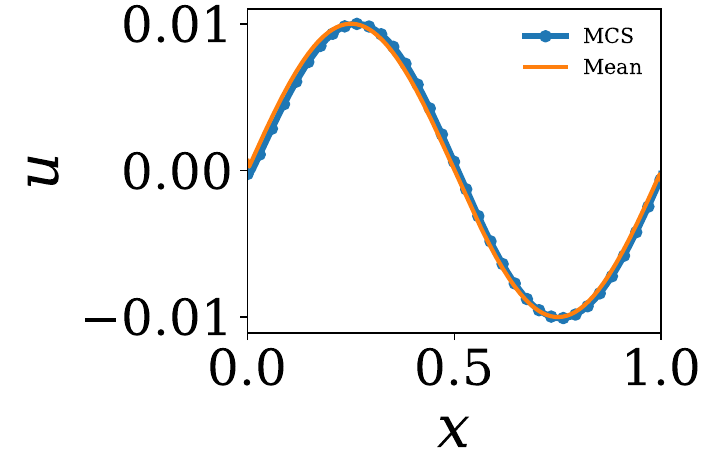} &
    \includegraphics[width=0.30\textwidth]{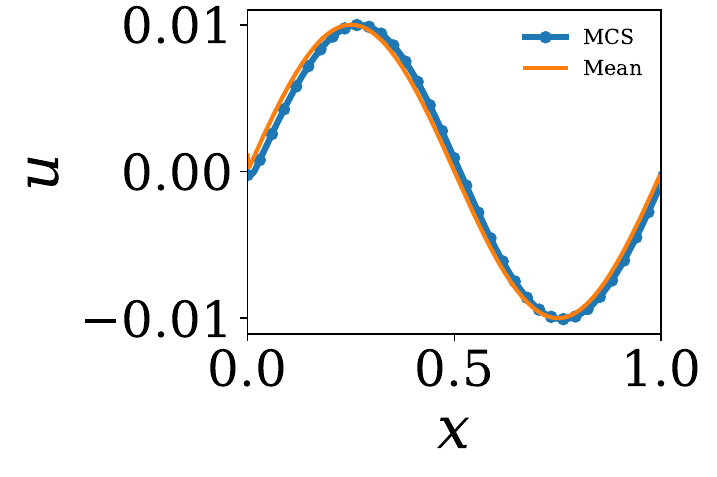} &
    \includegraphics[width=0.30\textwidth]{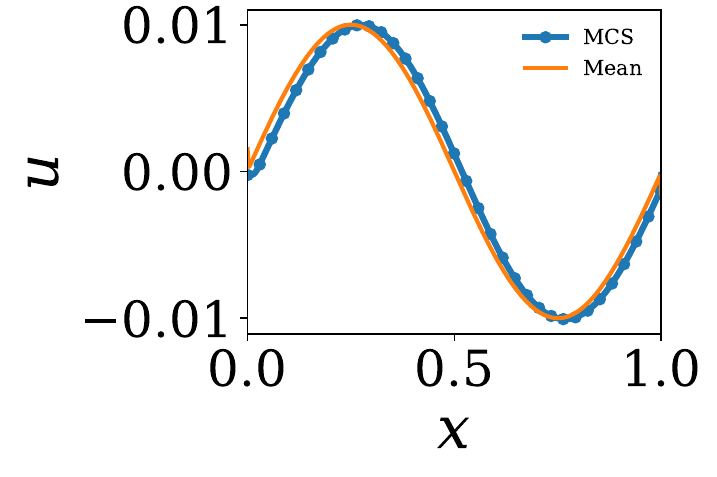}
  \end{tabular}

  \caption{%
  Mean response of the stochastic wave advection problem with the initial condition modeled as a Gaussian random field.
  The left column illustrates the spatio-temporal evolution of the mean field predicted using the S-SPH framework, while the corresponding MCS results are shown for comparison.
  Temporal histories at \(x = 0.30\) and \(0.70\) are shown in the right column, and spatial profiles extracted at \(T = 0.30s\), \(0.50s\), and \(0.70s\) are displayed in the bottom row.
  }
  \label{fig:ex1_mean_GRF}
\end{figure}

\begin{figure}[ht!]
  \centering

  \begin{tabular}{@{}c@{\hspace{1em}}c@{}}
    \includegraphics[width=0.65\textwidth]{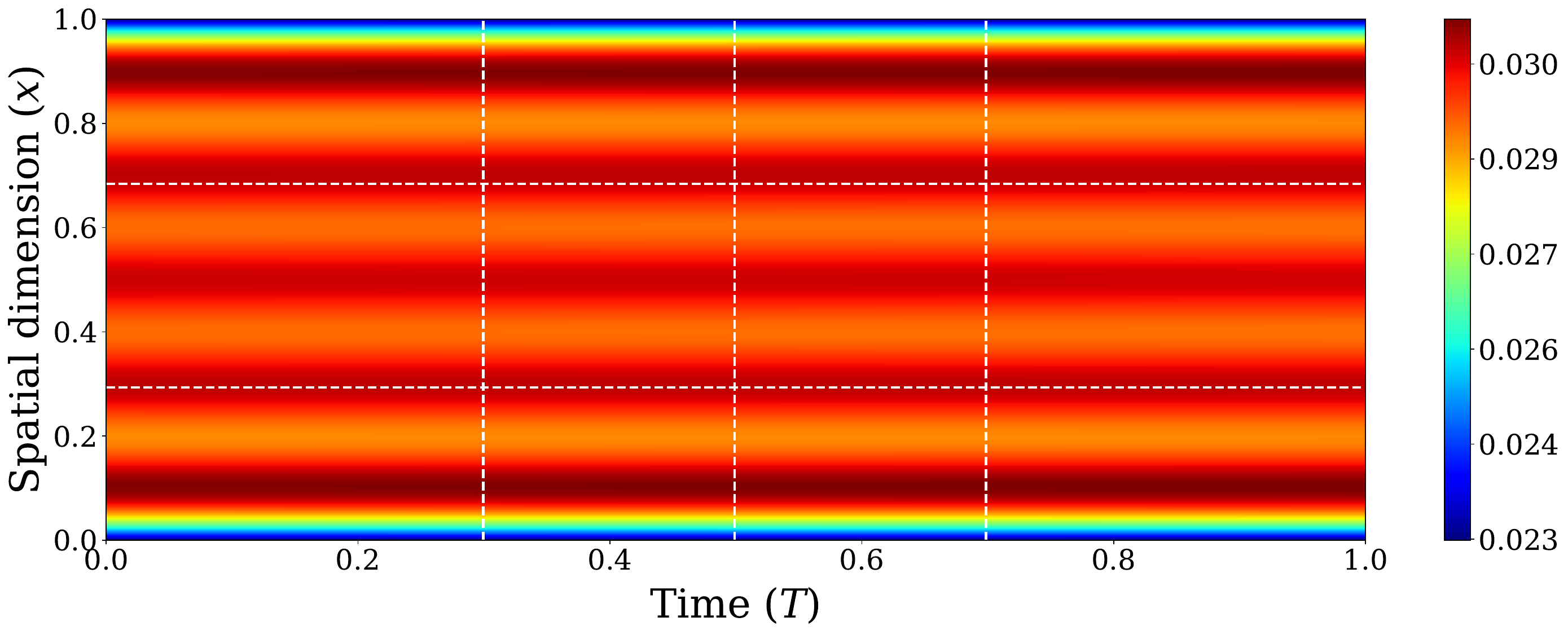} &
    \includegraphics[width=0.30\textwidth]{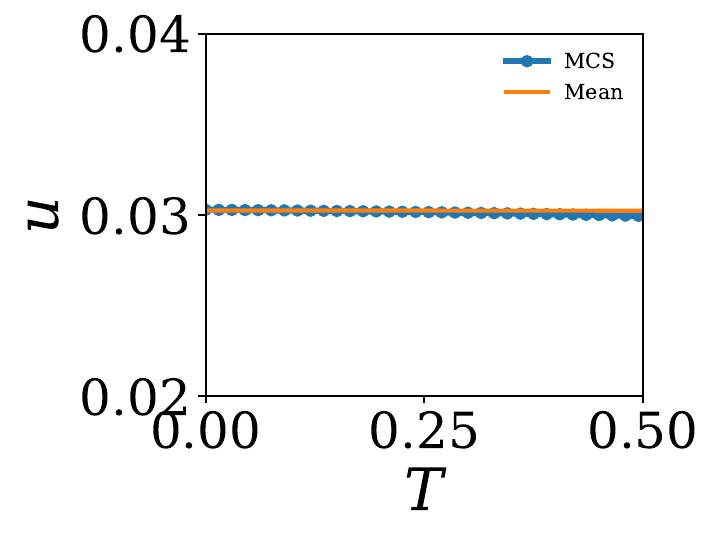}
  \end{tabular}

  \vspace{1em}

  \begin{tabular}{@{}c@{\hspace{1em}}c@{}}
    \includegraphics[width=0.65\textwidth]{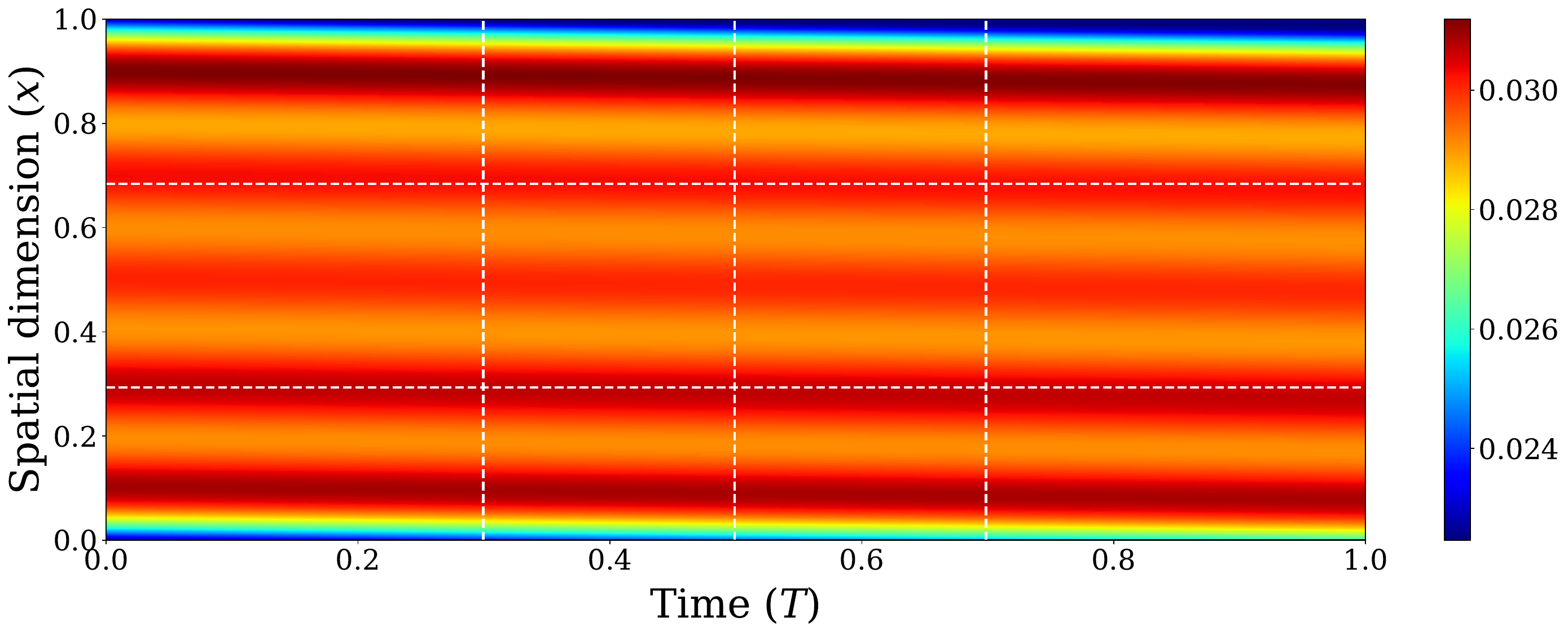} &
    \includegraphics[width=0.30\textwidth]{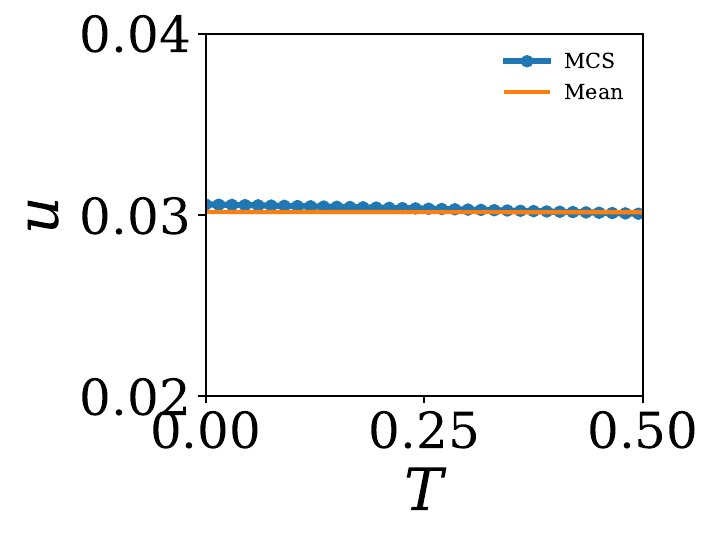}
  \end{tabular}

  \vspace{1em}

  \begin{tabular}{@{}c@{\hspace{1em}}c@{\hspace{1em}}c@{}}
    \includegraphics[width=0.30\textwidth]{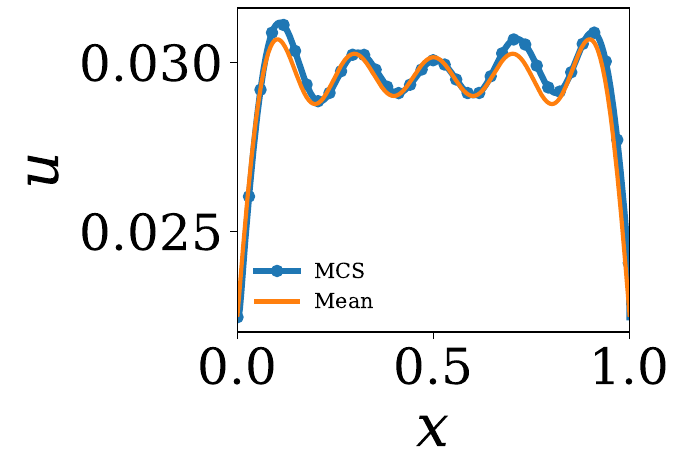} &
    \includegraphics[width=0.30\textwidth]{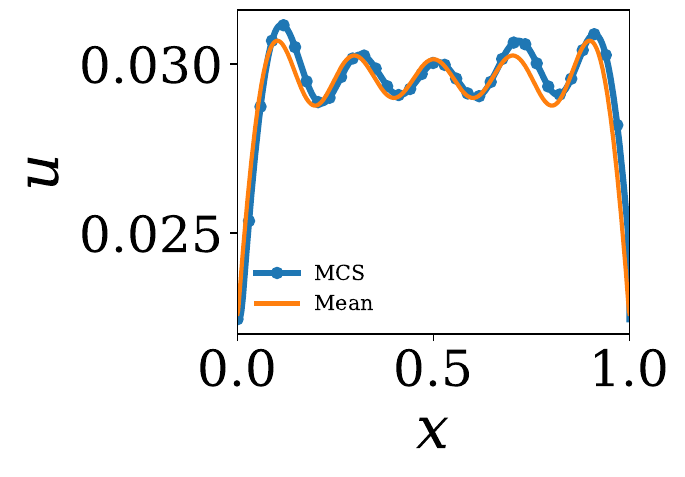} &
    \includegraphics[width=0.30\textwidth]{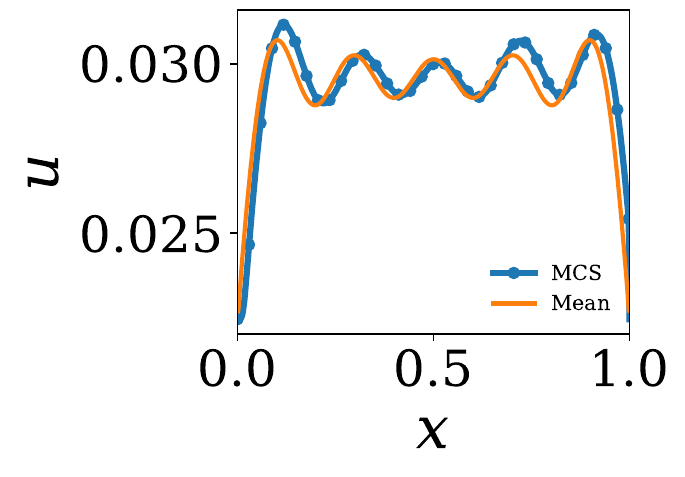}
  \end{tabular}

  \caption{%
  Standard deviation of the solution field for the stochastic wave advection problem with the initial condition modeled as a Gaussian random field.
  The top left column shows the spatio-temporal distribution of uncertainty predicted using the S-SPH framework, with corresponding MCS results provided for comparison.
  The right column shows temporal responses at \(x = 0.30\) and \(0.70\), while the bottom row presents spatial profiles at \(T = 0.30 s\), \(0.50 s\), and \(0.70 s\).
  The close agreement in uncertainty patterns across space and time confirms the consistency of the S-SPH estimates with MCS.
  }
  \label{fig:1_std_GRF}
\end{figure}

Fig.~\ref{fig:mean_std_u0_sin} presents the spatio-temporal contours of the response statistics for the first case, as computed using the proposed S-SPH method and MCS. Consistent with the previous examples, the S-SPH predictions exhibit near-identical agreement with the MCS reference, indicating high accuracy in capturing the response statistics.
Figs.~\ref{fig:ex1_mean_GRF} and~\ref{fig:1_std_GRF} show the corresponding mean and standard deviation fields for the second case, in which the initial condition is modeled as a Gaussian random field. In addition to the contour plots, pointwise comparisons of the response statistics at selected spatial and temporal locations are provided to further assess the predictive fidelity. Across all cases considered, the S-SPH results closely match the benchmark MCS solutions, demonstrating that the proposed framework can robustly and seamlessly accommodate random-field inputs.

\begin{figure}[ht!]
    \centering
    \begin{subfigure}[ht!]{0.48\textwidth}
        \centering
        \includegraphics[width=\textwidth]{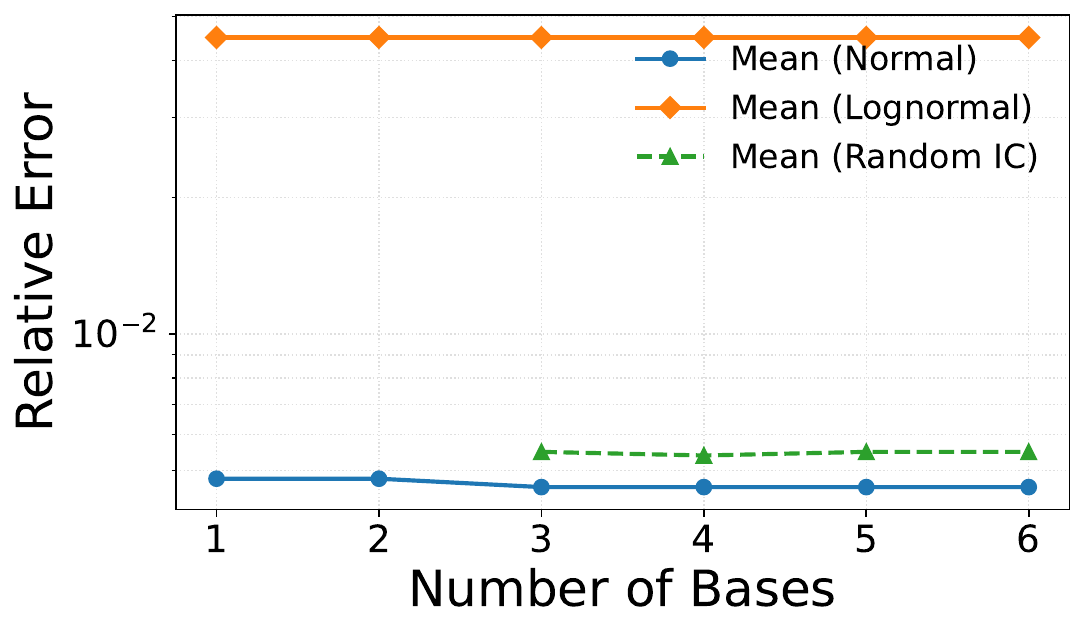}
        \caption{Mean relative $L_2$ error.}
        \label{fig:mean_error}
    \end{subfigure}\hfill
    \begin{subfigure}[ht!]{0.48\textwidth}
        \centering
        \includegraphics[width=\textwidth]{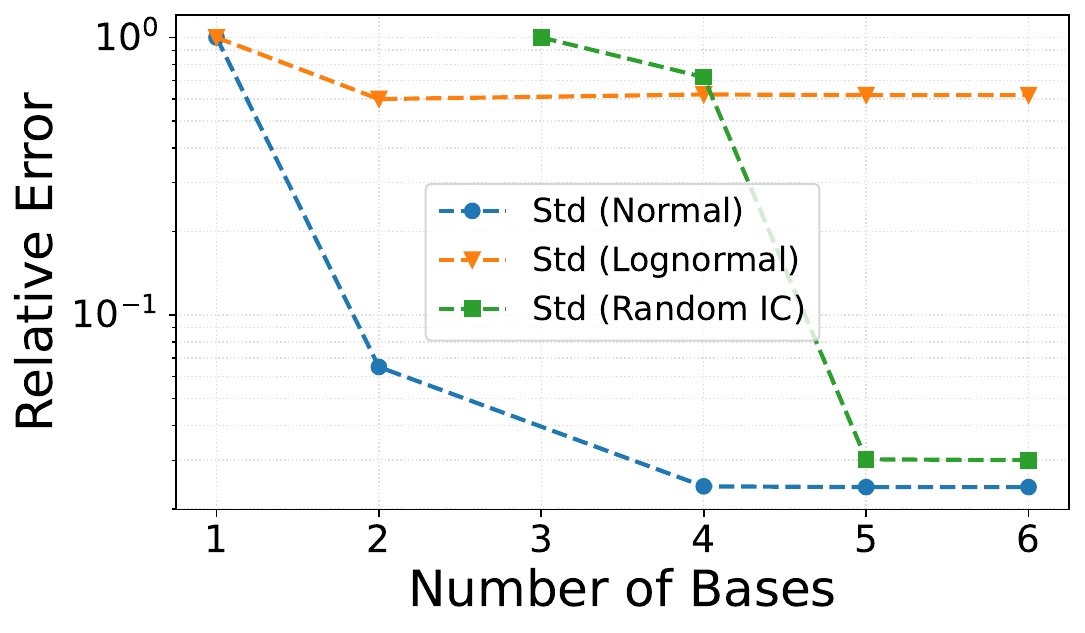}
        \caption{Std. of relative $L_2$ error.}
        \label{fig:std_error}
    \end{subfigure}
    \caption{Convergence behavior of the proposed S-SPH framework for Example~1, quantified in terms of the relative $L_2$ error $\epsilon_{\mathrm{rel}}$ with respect to the MCS benchmark. Figures (a) and (b) report the mean and standard deviation of the relative $L_2$ error. It is plotted as a function of the number of orthogonal basis functions $P$ used in the polynomial chaos expansion.}
    \label{fig:error_vs_bases}
\end{figure}

Finally,
a convergence study is performed to assess the influence of the number of basis functions on the accuracy of the proposed S-SPH framework. To this end, Fig.~\ref{fig:error_vs_bases} reports the relative error as a function of the number of retained basis functions. As expected, the error decreases monotonically with increasing basis dimensionality, indicating systematic convergence of the method.
For all cases considered, the mean response converges rapidly, with satisfactory accuracy achieved using three basis functions (third-order approximation). In contrast, accurate estimation of the standard deviation requires up to five basis functions to attain convergence. This behavior is consistent with the well-known observation that higher-order statistics are more sensitive to approximation errors and therefore demand richer representations than first-order moments.

\subsection{Example 2: 1D inviscid Burgers' Equation}\label{example_2}
As a second test problem, we consider the inviscid Burgers’ equation, a canonical nonlinear hyperbolic partial differential equation widely used to model shock formation, nonlinear wave propagation, and traffic flow. In its non-conservative form, the governing equation is given by
\begin{equation}
\frac{\partial u(x,t,\omega)}{\partial t} 
+ u(x,t,\omega)\,\frac{\partial u(x,t,\omega)}{\partial x} = 0,
\quad t \in [0,T], \; x \in [0,1],
\end{equation}
subject to periodic boundary conditions,
\begin{equation}
u(-1,t) = u(1,t), \quad t \in [0,T],
\end{equation}
and the initial condition
\begin{equation}
u(x,0) = u_0(x,\omega), \quad x \in [0,1].
\end{equation}
Here, \( u(x,t,\omega) \) denotes a stochastic velocity field defined over the probability space \( \omega \in \Omega \). 

Enforcing periodic boundary conditions constitutes a key numerical challenge in S-SPH. We impose periodicity by modifying the particle neighbor-search algorithm. Specifically, particles located near one boundary of the computational domain (e.g., \( x = 0 \)) are treated as neighbors of particles located near the opposite boundary (e.g., \( x = 1 \)) during neighbor list construction. This wrap-around strategy effectively renders the domain periodic, ensuring continuity of particle interactions and information transfer across boundaries. Notably, this approach enforces periodicity without introducing ghost or mirror particles, thereby preserving computational efficiency and consistency with the particle-based formulation.

Uncertainty is introduced through the initial condition, for which two stochastic representations are considered. In the first case, analogous to the previous example, the initial condition is parameterized as
\begin{equation}
u_0(x) = \alpha \sin(\beta x),
\end{equation}
where the amplitude \( \alpha \sim \mathcal{N}(0.25,\,0.1) \) and the wavenumber \( \beta \sim \mathcal{N}(2\pi,\,0.1) \) are modeled as independent Gaussian random variables, thereby inducing variability in both the magnitude and spatial frequency of the initial profile.
In the second case, the initial condition is modeled as a Gaussian random field with a mean function
\begin{equation}
\mu_{u_0}(x) = 0.01 \sin(2\pi x),
\end{equation}
and covariance kernel
\begin{equation}\label{eq:covariance_kernel_rf2}
k_{u_0}(x,x') = \sigma_{u_0}^2 
\exp\!\left(-\frac{\|x - x'\|_2^2}{l_{u_0}^2}\right),
\end{equation}
where \( \sigma_{u_0}^2 \) denotes the process variance and \( l_{u_0} \) is the characteristic length scale. For this study, the parameters are set to 
\( \sigma_{u_0}^2 = 0.001 \) and \( l_{u_0} = 0.01 \).
To obtain a finite-dimensional representation of the random field, the Karhunen--Lo\`eve (KL) expansion described in Eq.~\eqref{eq:kle} is employed.

\begin{figure}[ht!]
  \centering

  \begin{tabular}{@{}c@{\hspace{1em}}c@{}}
    \includegraphics[width=0.65\textwidth]{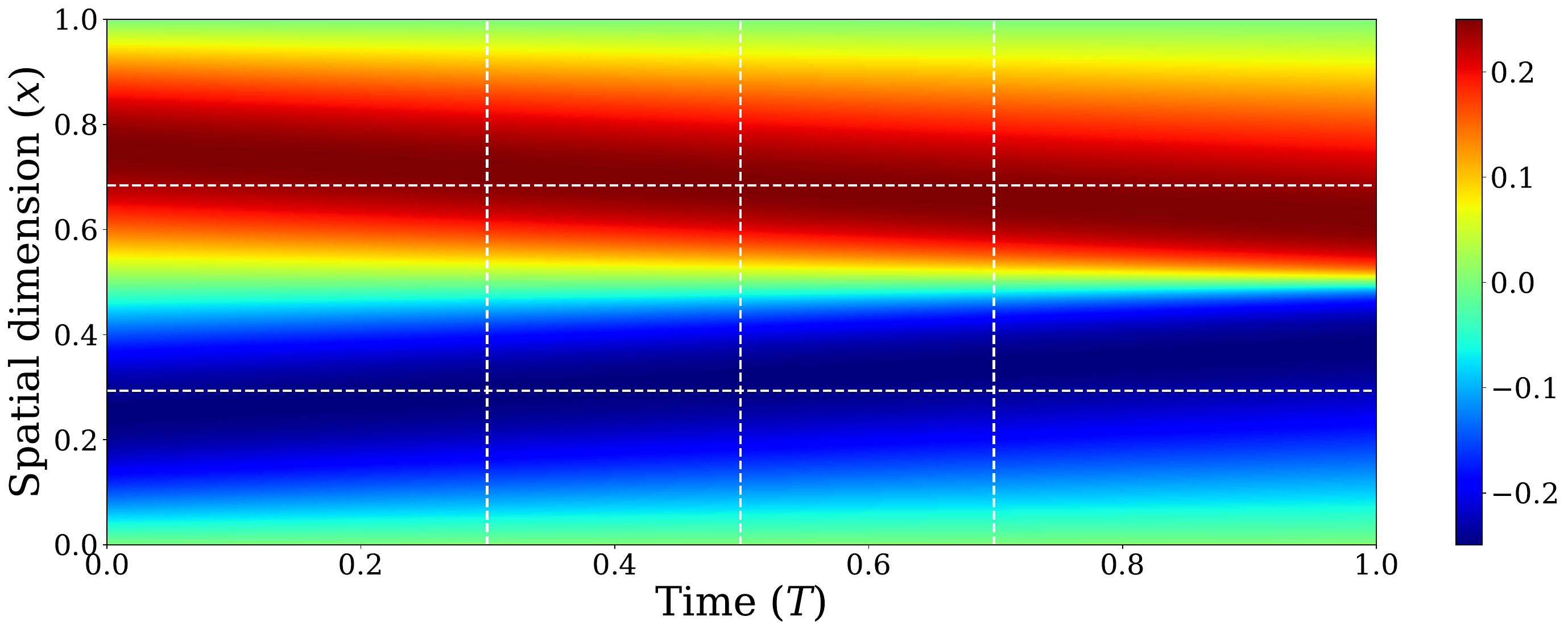} &
    \includegraphics[width=0.30\textwidth]{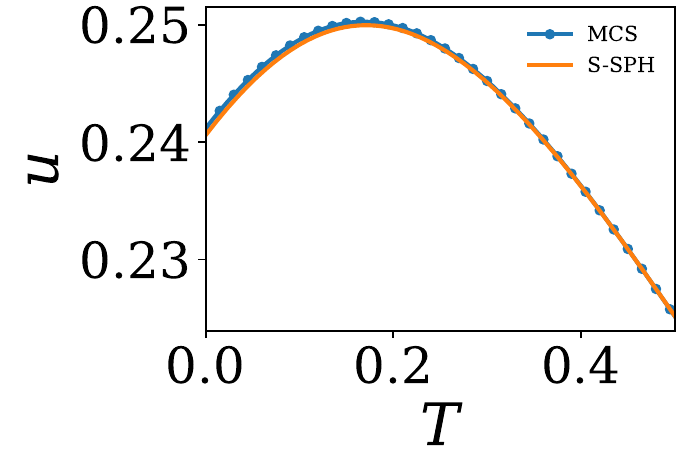}
  \end{tabular}

  \vspace{1em}

  \begin{tabular}{@{}c@{\hspace{1em}}c@{}}
    \includegraphics[width=0.65\textwidth]{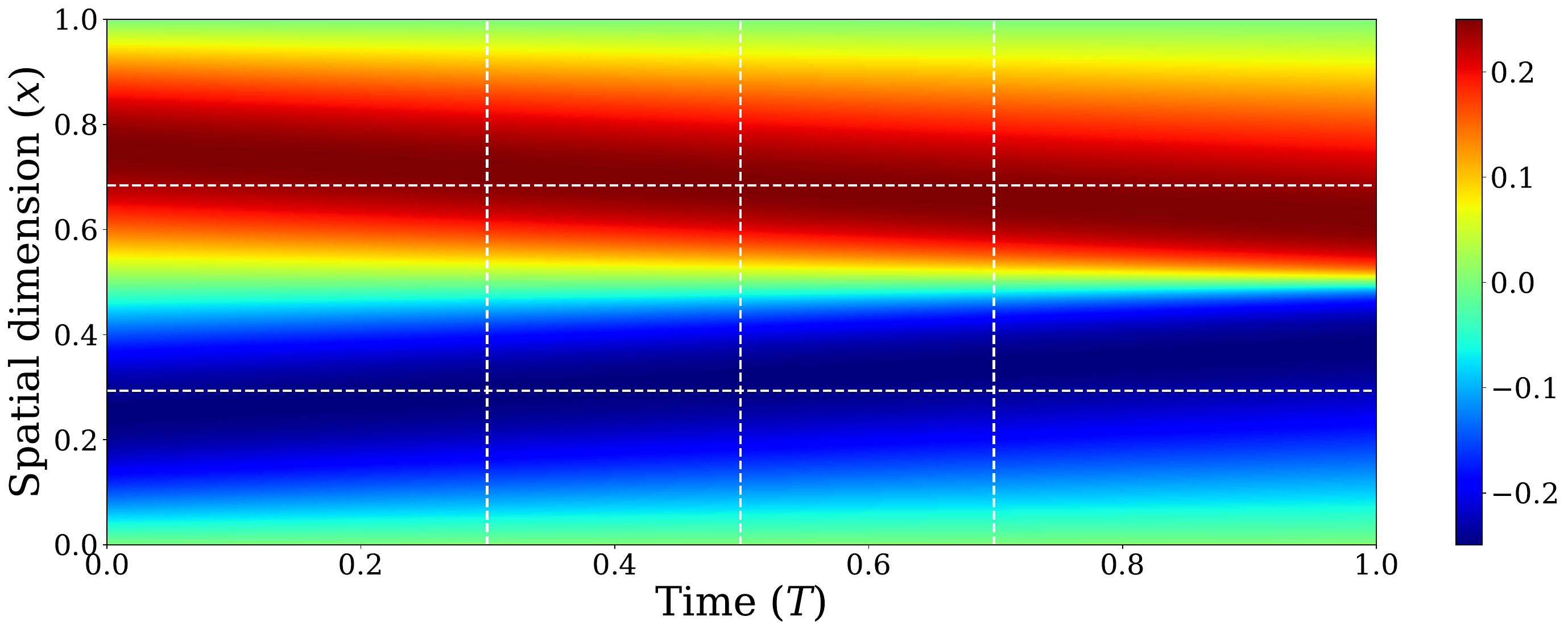} &
    \includegraphics[width=0.30\textwidth]{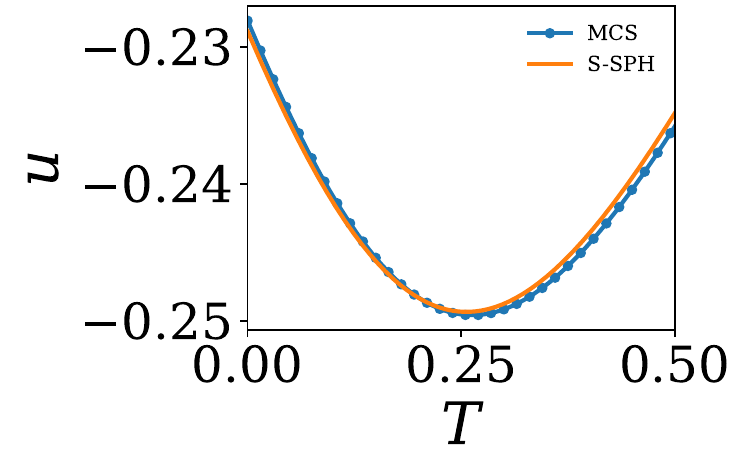}
  \end{tabular}

  \vspace{1em}

  \begin{tabular}{@{}c@{\hspace{1em}}c@{\hspace{1em}}c@{}}
    \includegraphics[width=0.30\textwidth]{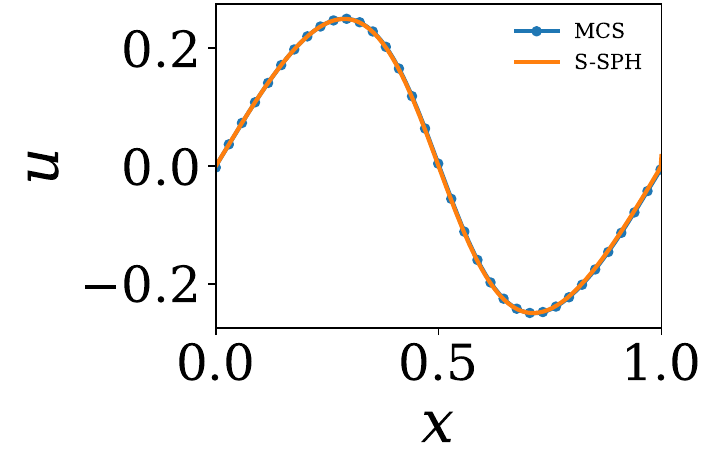} &
    \includegraphics[width=0.30\textwidth]{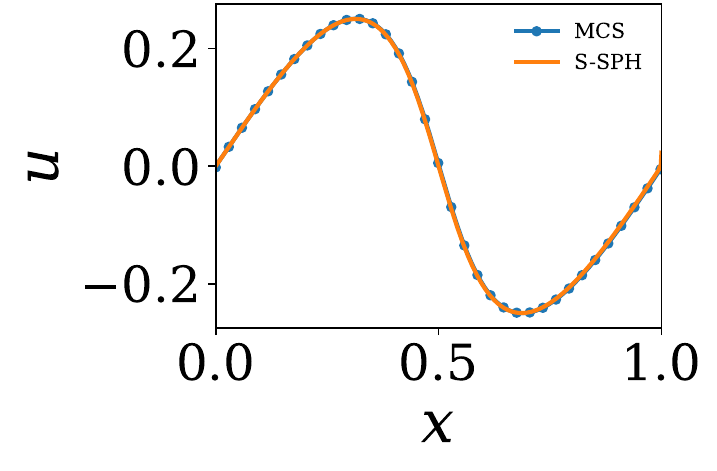} &
    \includegraphics[width=0.30\textwidth]{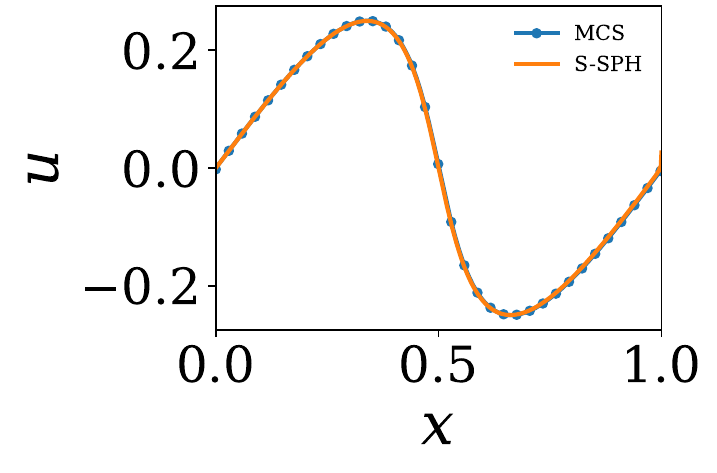}
  \end{tabular}

  \caption{%
  Mean response for Case~1 with random initial conditions, comparing the S-SPH framework with MCS.
  The left column illustrates the spatio-temporal evolution of the mean solution predicted using S-SPH, with the corresponding MCS results shown for comparison.
  Temporal histories at \(x = 0.30\) and \(0.70\) are shown in the right column, and spatial profiles extracted at \(T = 0.30s\), \(0.50s\), and \(0.70s\) are shown in the bottom row.
  }
  \label{fig:ex2_case1_mean}
\end{figure}

\begin{figure}[htbp!]
  \centering

  \begin{tabular}{@{}c@{\hspace{1em}}c@{}}
    \includegraphics[width=0.65\textwidth]{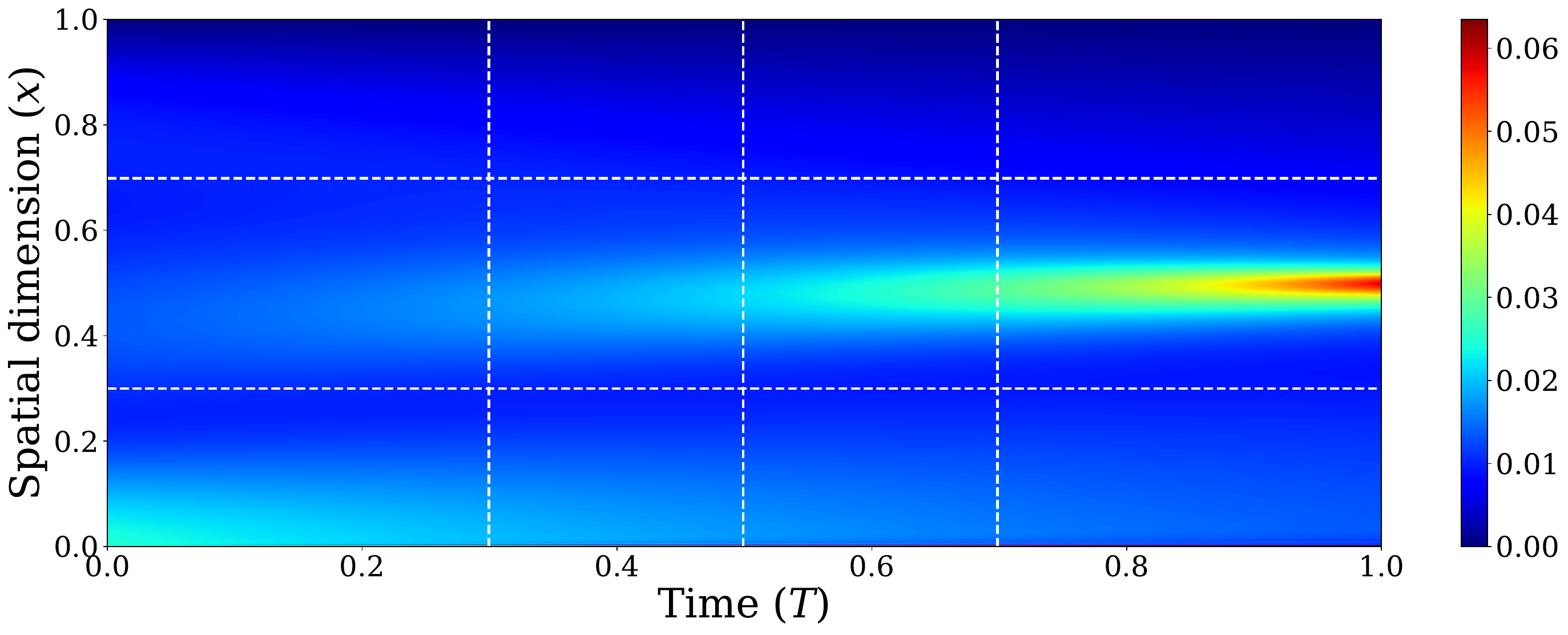} &
    \includegraphics[width=0.30\textwidth]{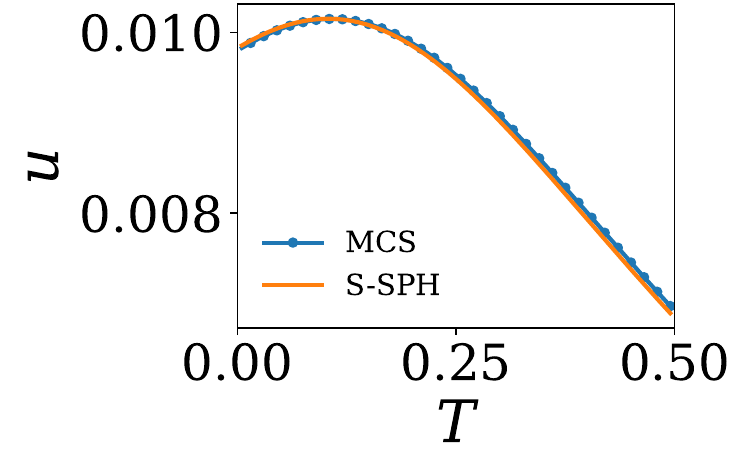}
  \end{tabular}

  \vspace{1em}

  \begin{tabular}{@{}c@{\hspace{1em}}c@{}}
    \includegraphics[width=0.65\textwidth]{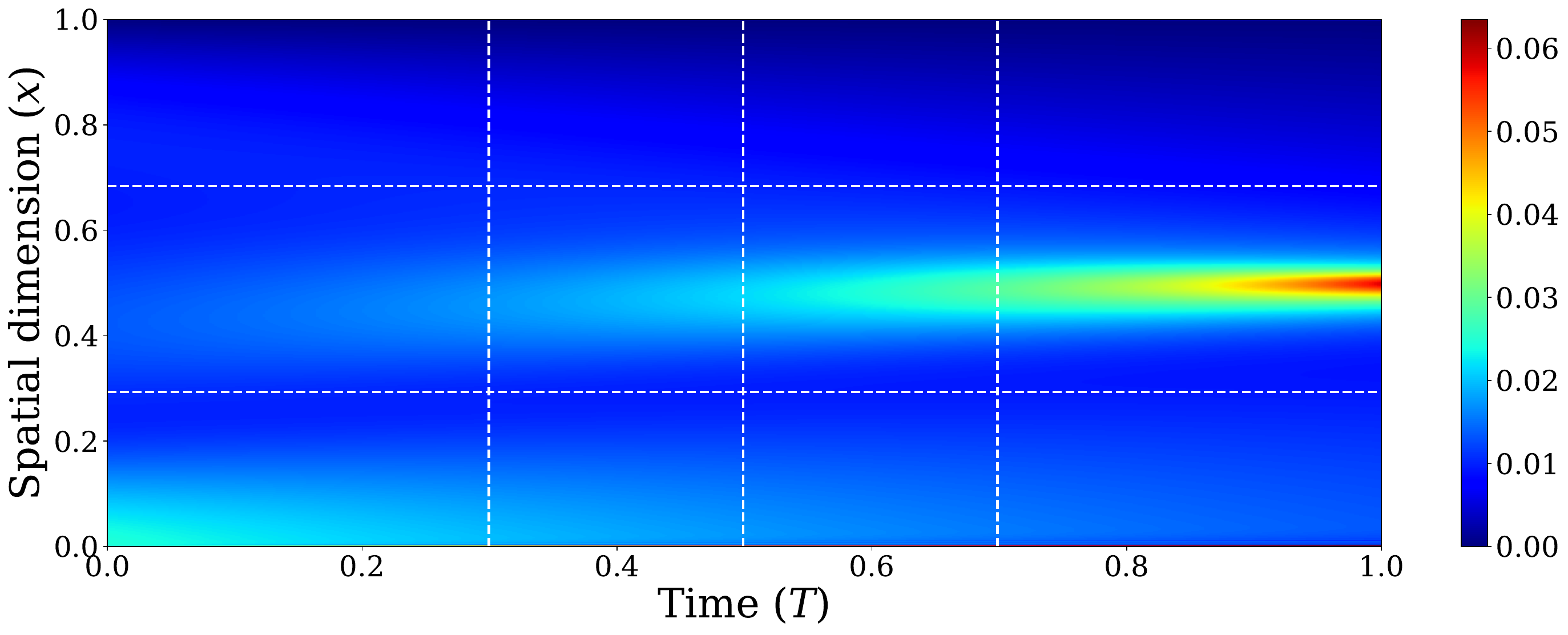} &
    \includegraphics[width=0.30\textwidth]{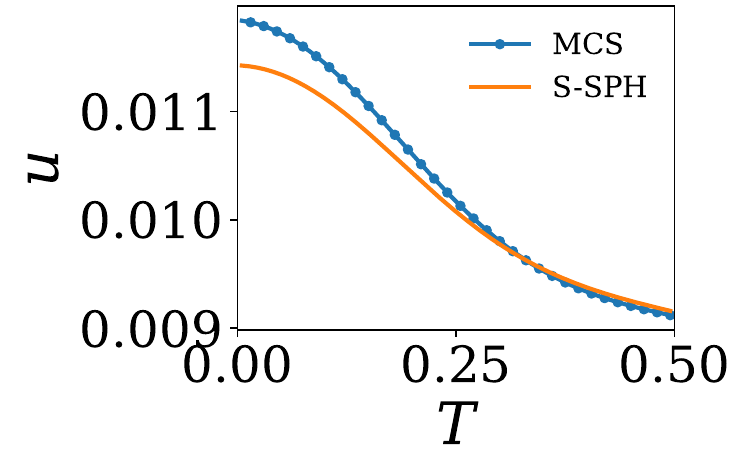}
  \end{tabular}

  \vspace{1em}

  \begin{tabular}{@{}c@{\hspace{1em}}c@{\hspace{1em}}c@{}}
    \includegraphics[width=0.30\textwidth]{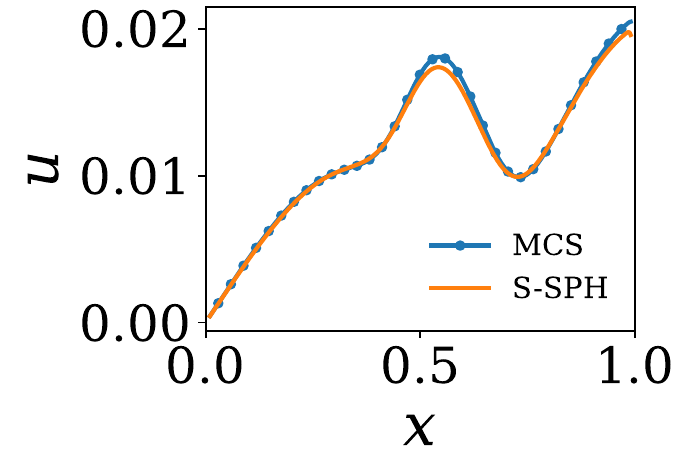} &
    \includegraphics[width=0.30\textwidth]{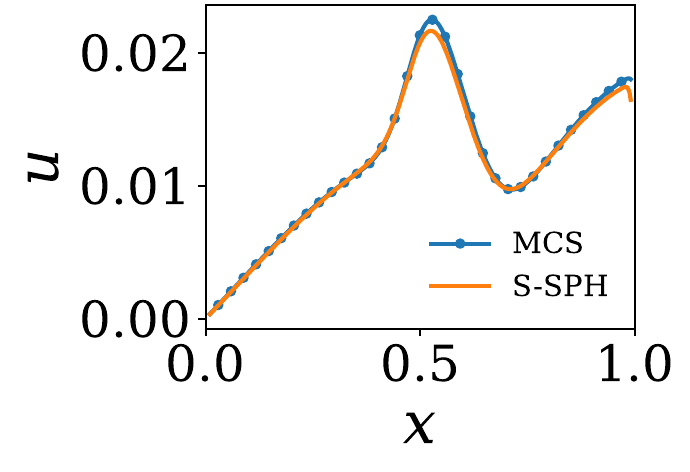} &
    \includegraphics[width=0.30\textwidth]{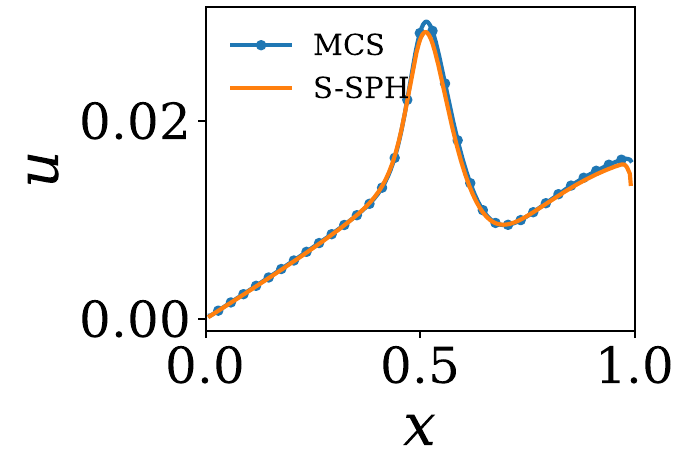}
  \end{tabular}

  \caption{%
  Standard deviation of the solution field for Case~1 with random initial conditions, comparing results obtained using the S-SPH framework and MCS.
  The top left column depicts the spatio-temporal evolution of uncertainty predicted by S-SPH, alongside the corresponding MCS reference.
  The bottom row depicts spatial snapshots at selected times, and the right column reports temporal variations. Spatial profiles are extracted at \(T = 0.30s\), \(0.50s\), and \(0.70s\), and temporal responses are evaluated at \(x = 0.30\) and \(0.70\).
  }
  \label{fig:ex2_case1_std}
\end{figure}

\begin{figure}[ht!]
  \centering

  \begin{tabular}{@{}c@{\hspace{1em}}c@{}}
    \subcaptionbox{Mean S-SPH\label{fig:mean_ssph_ex2}}{%
      \includegraphics[width=0.45\textwidth]{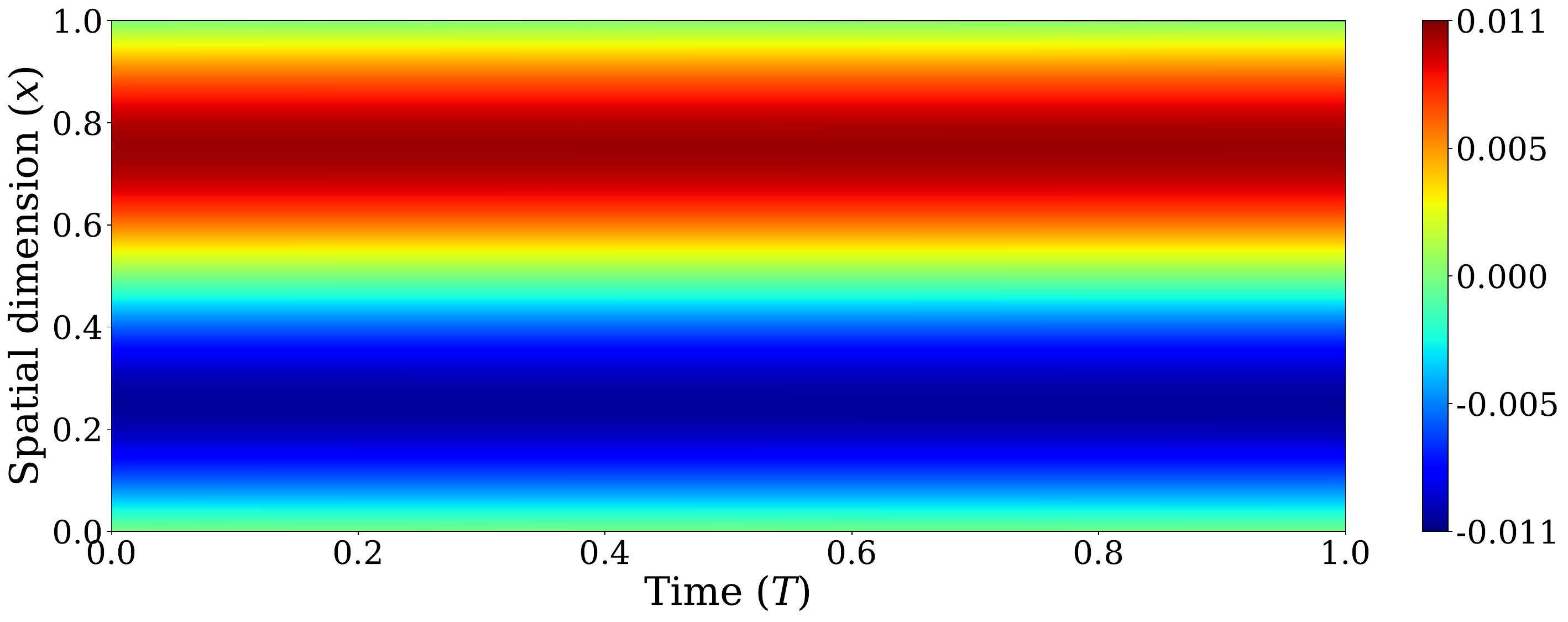}} &
    \subcaptionbox{Mean MCS\label{fig:mean_mcs_ex2}}{%
      \includegraphics[width=0.45\textwidth]{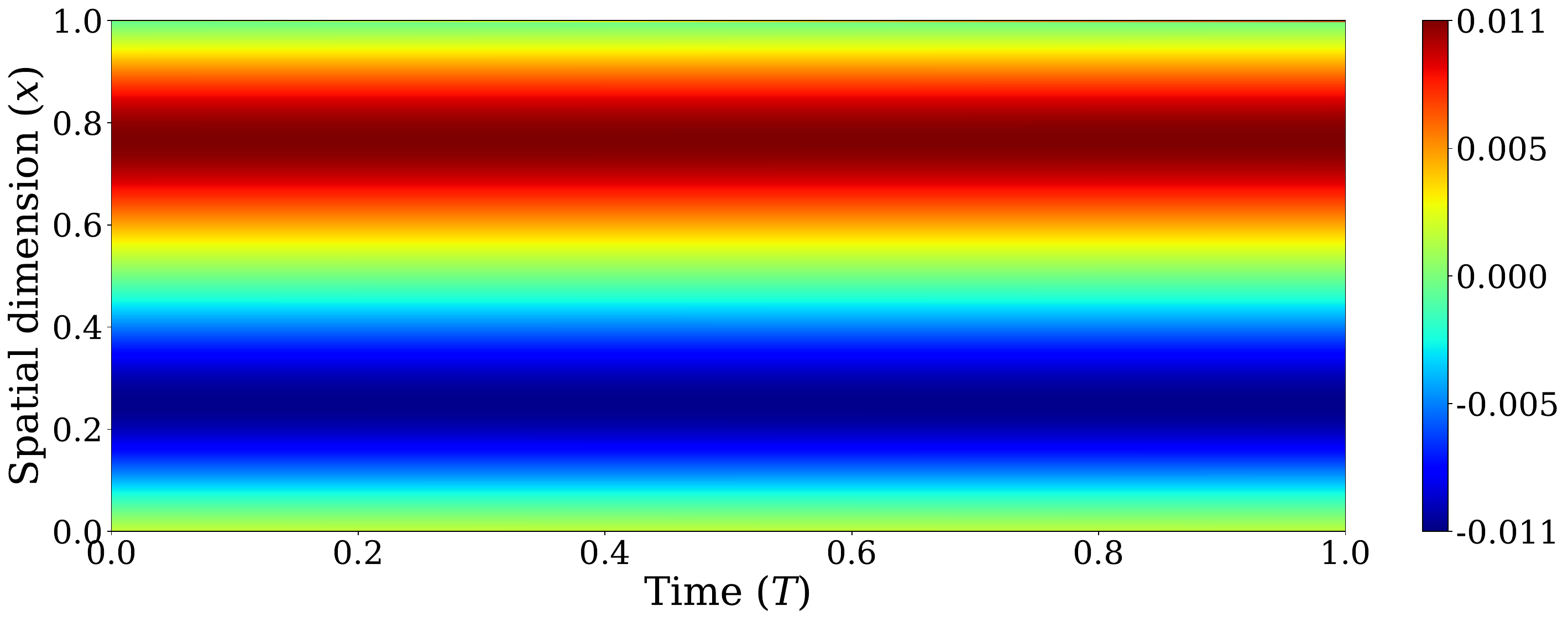}}
  \end{tabular}

  \vspace{1em}

  \begin{tabular}{@{}c@{\hspace{1em}}c@{}}
    \subcaptionbox{Std S-SPH\label{fig:std_ssph_ex2}}{%
      \includegraphics[width=0.45\textwidth]{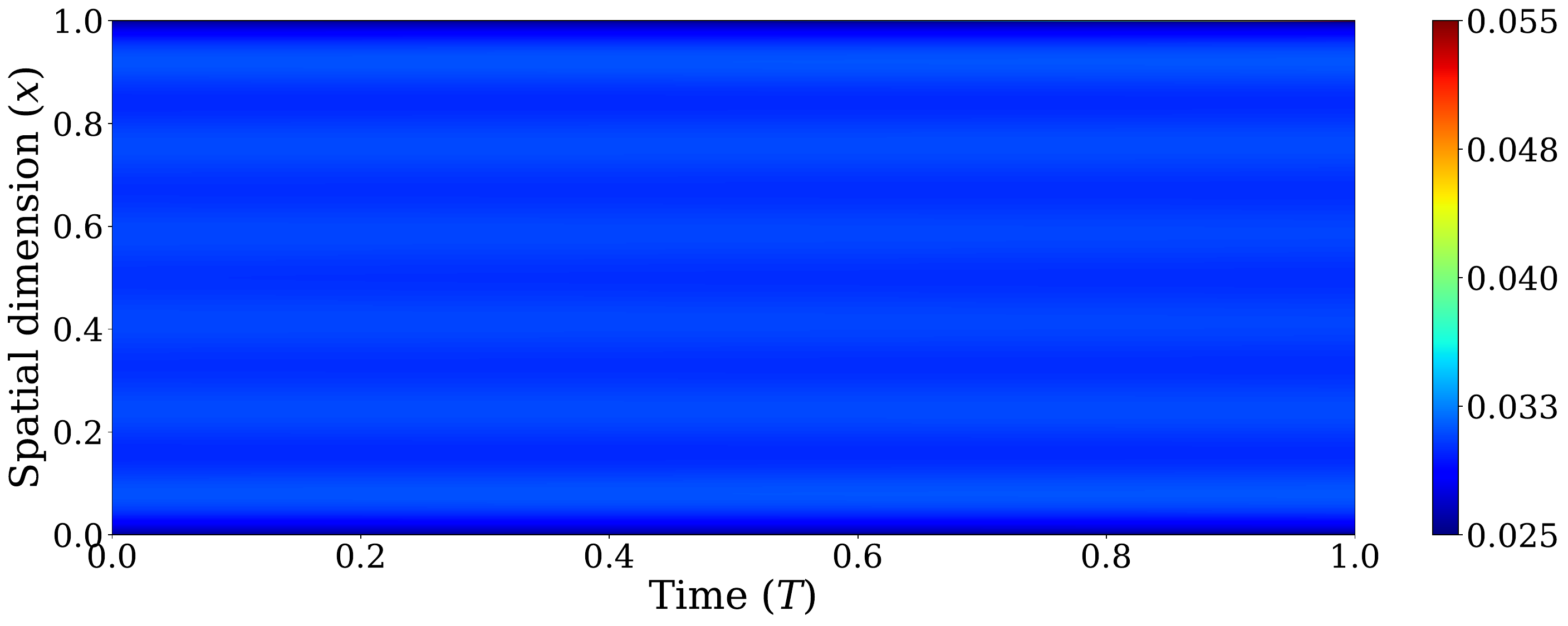}} &
    \subcaptionbox{Std MCS\label{fig:std_mcs_ex2}}{%
      \includegraphics[width=0.45\textwidth]{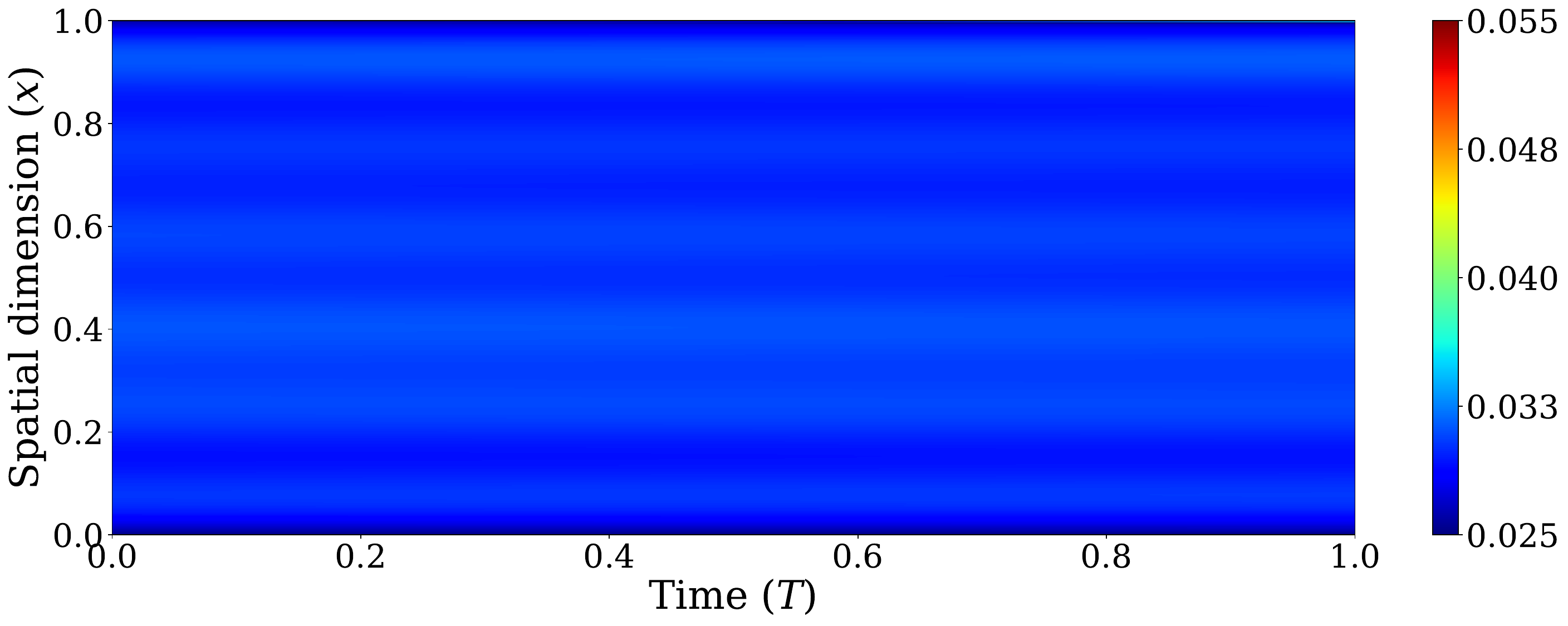}}
  \end{tabular}

  \caption{Mean and standard deviation contours for Example~2, Case~2 of the stochastic problem.
  Subfigures (a) and (b) show the spatio-temporal evolution of the mean response obtained using the S-SPH framework and MCS, respectively.
  Subfigures (c) and (d) present the corresponding standard deviation fields.
  The close agreement between S-SPH and MCS in both the mean and uncertainty measures demonstrates the consistency of the proposed approach for this test case.
}
  \label{fig:mean_std_ex2_c2}
\end{figure}

Figs.~\ref{fig:ex2_case1_mean} and~\ref{fig:ex2_case1_std} present the spatio-temporal contours of the response mean and standard deviation for Case~1. The S-SPH predictions show an almost identical agreement with the MCS results, accurately capturing both first- and second-order response statistics. To further assess the predictive fidelity of the proposed method, pointwise comparisons of the response statistics at selected spatial and temporal locations are also performed. While the mean response obtained using S-SPH matches the MCS results exactly, a slight deviation is observed in the standard deviation at one spatial location. This discrepancy is minimal, remains localized, and does not affect the overall agreement between the two approaches.
Fig.~\ref{fig:mean_std_ex2_c2} shows the corresponding mean and standard deviation responses for Case~2. In this case as well, the response statistics predicted by S-SPH exhibit close agreement with the MCS reference across the domain, further demonstrating the robustness of the proposed framework in capturing uncertainty propagation for both parametric and random-field inputs.

Fig. \ref{fig:mean_std_ex2_c2} shows the mean and the standard deviation of the response for the second case. In this case also, the response statistics obtained using the proposed approach matches quite well with the MCS result.

\subsection{Example 3: Two-Dimensional Burgers’ Equation}\label{example_3}

In the third example, we extend the one-dimensional studies to the two-dimensional viscous Burgers’ equation, which serves as a canonical nonlinear model for examining the interplay between advection and diffusion in the presence of nonlinear transport. Owing to its nonlinear convective term and diffusive regularization, this problem provides a stringent benchmark for assessing the robustness and scalability of the proposed S-SPH framework in higher dimensions.
The governing equations are given by
\begin{equation}
\frac{\partial \mathbf{u}}{\partial t}
+ \mathbf{u}\cdot\nabla \mathbf{u}
- \nu\,\Delta \mathbf{u}
= \mathbf{0},
\quad (x,y)\in(0,1)\times(0,1), \; t>0,
\end{equation}
subject to homogeneous Dirichlet boundary conditions,
\begin{equation}
\mathbf{u}(0,y,t) = \mathbf{u}(1,y,t) = \mathbf{0}, 
\qquad
\mathbf{u}(x,0,t) = \mathbf{u}(x,1,t) = \mathbf{0},
\quad t>0.
\end{equation}
Here, $\mathbf{u}=(u,v)$ denotes the two-dimensional velocity field, and $\nu$ is the kinematic viscosity.
All stochastic simulations are performed using the proposed S-SPH framework with a basis of order $P=3$. Temporal integration is carried out using a time step of $\Delta t = 0.001$ over a total simulation time of $T = 0.5$, resulting in $N = T/\Delta t$ time steps. The spatial domain is discretized using a uniform particle grid with resolution $J = 128$ in each direction, corresponding to $\Delta x = 1/J$. The smoothing length is set to $h = 1.6\,\Delta x$, with an influence radius of $r = 2h$. In all cases, MCS with $5000$ samples is used as the reference solution for validation.

We consider two stochastic configurations to investigate the performance of the S-SPH method under different sources of uncertainty. In the first configuration, uncertainty is introduced through the initial velocity field, which is constructed from a truncated two-dimensional Fourier expansion. Specifically, the initial condition is defined as
\begin{equation}
u(x,y,0) = \frac{2\,w(x,y)}{\max_{x,y}|w(x,y)|},
\end{equation}
where
\begin{equation}
w(x,y) =
\sum_{i=-4}^{4}\sum_{j=-4}^{4}
\Bigl[
a_{ij}\sin\bigl(2\pi(i x + j y)\bigr)
+
b_{ij}\cos\bigl(2\pi(i x + j y)\bigr)
\Bigr].
\end{equation}
The Fourier coefficients $a_{ij}$ and $b_{ij}$ are assumed to be independent and identically distributed Gaussian random variables, $a_{ij},\,b_{ij} \sim \mathcal{N}(0,1)$. The normalization ensures bounded initial velocities while preserving the stochastic spatial structure. 
In the second stochastic configuration, uncertainty is introduced through the viscosity coefficient, which is modeled as a spatially varying random field. The viscosity is represented using a Karhunen--Lo\`eve expansion,
\begin{equation}
\nu(x,y,\omega) = \bar{\nu}
+ \sum_{k=1}^{N_{\mathrm{KL}}}
\sqrt{\lambda_k}\,\xi_k(\omega)\,\phi_k(x,y),
\end{equation}
where $\bar{\nu}=0.05$ denotes the mean viscosity, $\{(\lambda_k,\phi_k)\}$ are the eigenpairs of the prescribed covariance kernel, and $\xi_k \sim \mathcal{N}(0,1)$ are independent standard Gaussian random variables. This representation provides a low-dimensional parameterization of the stochastic viscosity field, enabling efficient uncertainty propagation within the S-SPH framework. Notably, this case introduces multiplicative uncertainty directly into the diffusion operator, making it a particularly challenging test for stochastic solvers.

\begin{figure}[htbp!]
    \centering
    \textbf{} \par\medskip
    \begin{subfigure}[b]{\textwidth}
        \centering
        \includegraphics[width=\textwidth]{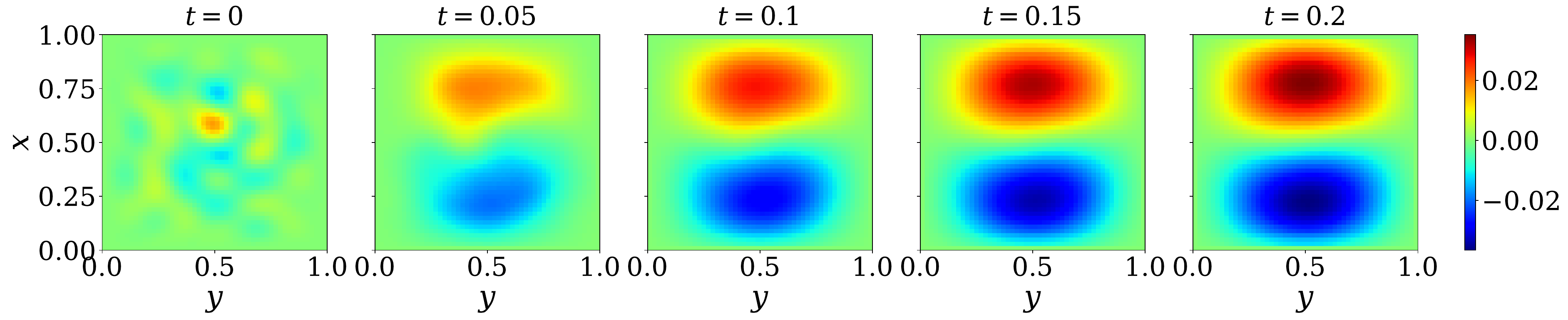}
        \caption{Mean (S-SPH)}
    \end{subfigure}
    \begin{subfigure}[b]{\textwidth}
        \centering
        \includegraphics[width=\textwidth]{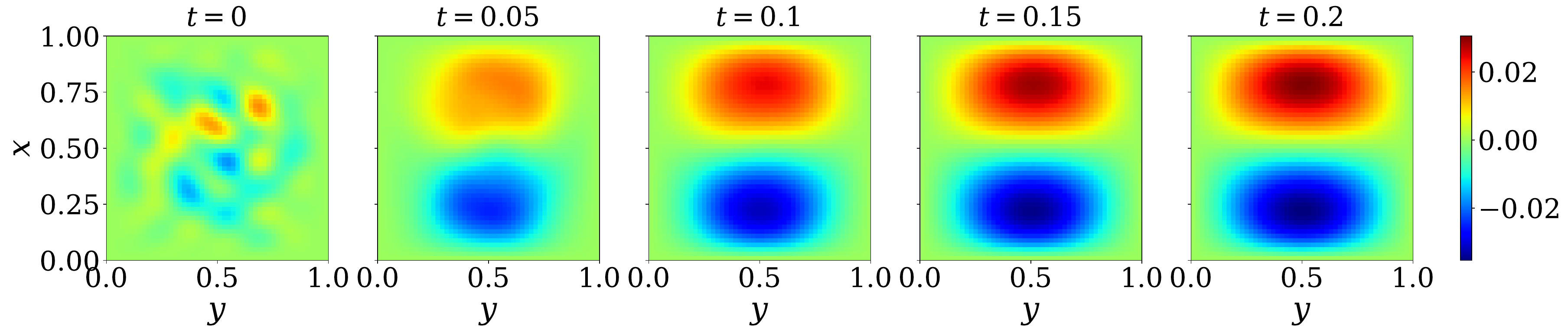}
        \caption{Mean (MCS)}
    \end{subfigure}
    \begin{subfigure}[b]{\textwidth}
        \centering
        \includegraphics[width=\textwidth]{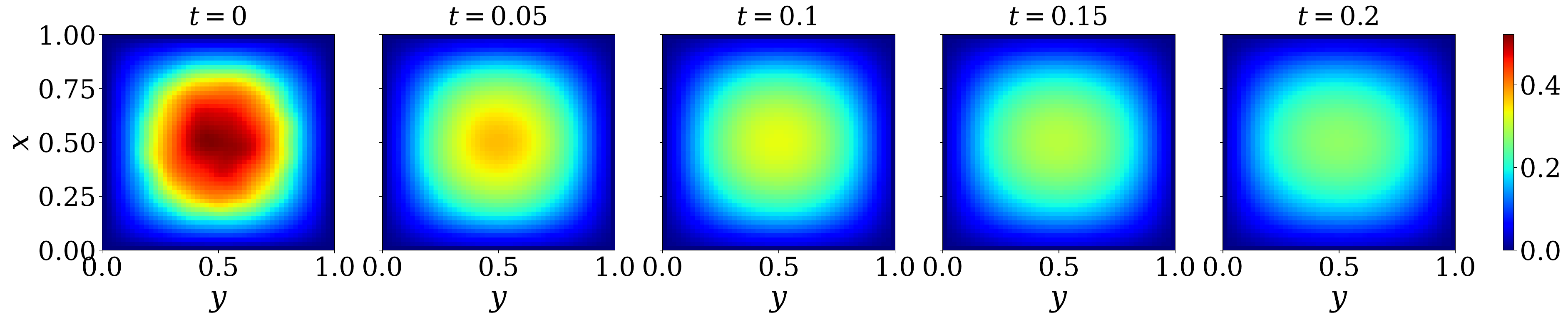}
        \caption{Mean (S-SPH)}
    \end{subfigure}
    \begin{subfigure}[b]{\textwidth}
        \centering
        \includegraphics[width=\textwidth]{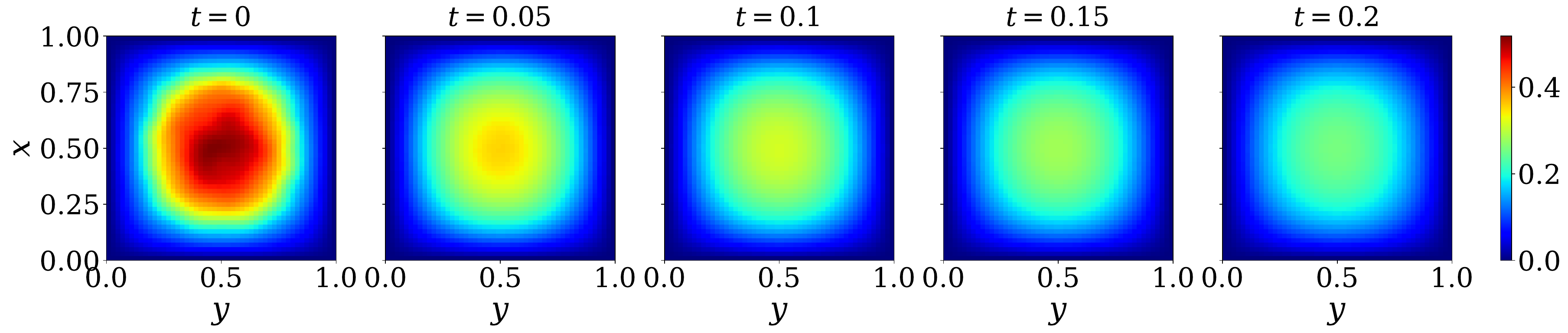}
        \caption{Std (MCS)}
    \end{subfigure}
    \caption{Time-resolved contour figures for the 2-D stochastic Burgers’ equation with the initial condition prescribed as a random field (Case~1). 
    The first two rows show the evolution of the mean velocity field obtained using S-SPH and MCS, respectively, at selected time instants. 
    The bottom two rows present the corresponding standard deviation fields, highlighting the spatial distribution and temporal decay of uncertainty.
    The close visual correspondence between S-SPH and MCS across all subfigures indicates that the proposed method accurately captures both the evolving flow structure and its associated uncertainty.
}
    \label{fig:ex3_case1}
\end{figure}

\begin{figure}[htbp]
    \centering
    \textbf{} \par\medskip
    \begin{subfigure}[b]{\textwidth}
        \centering
        \includegraphics[width=\textwidth]{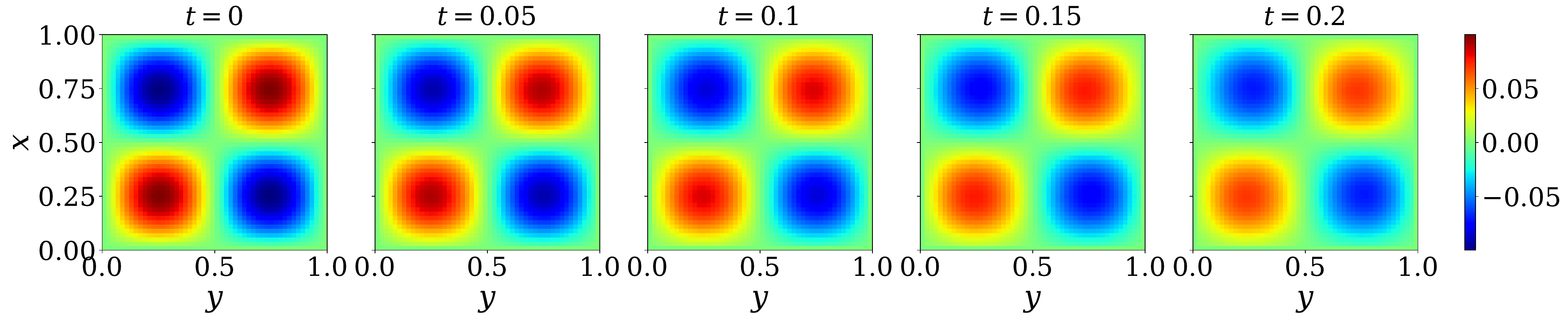}
        \caption{Mean (S-SPH)}
    \end{subfigure}
    \begin{subfigure}[b]{\textwidth}
        \centering
        \includegraphics[width=\textwidth]{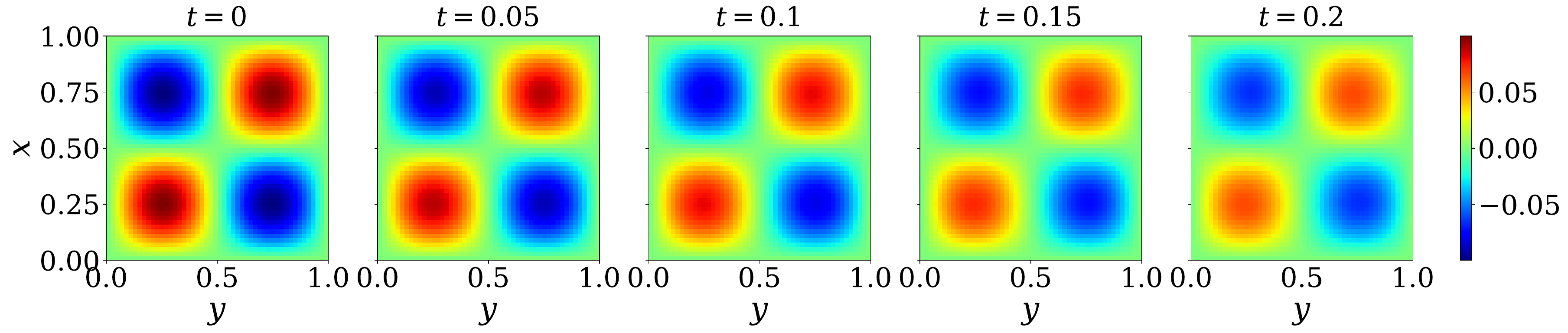}
        \caption{Mean (MCS)}
    \end{subfigure}
    \begin{subfigure}[b]{\textwidth}
        \centering
        \includegraphics[width=\textwidth]{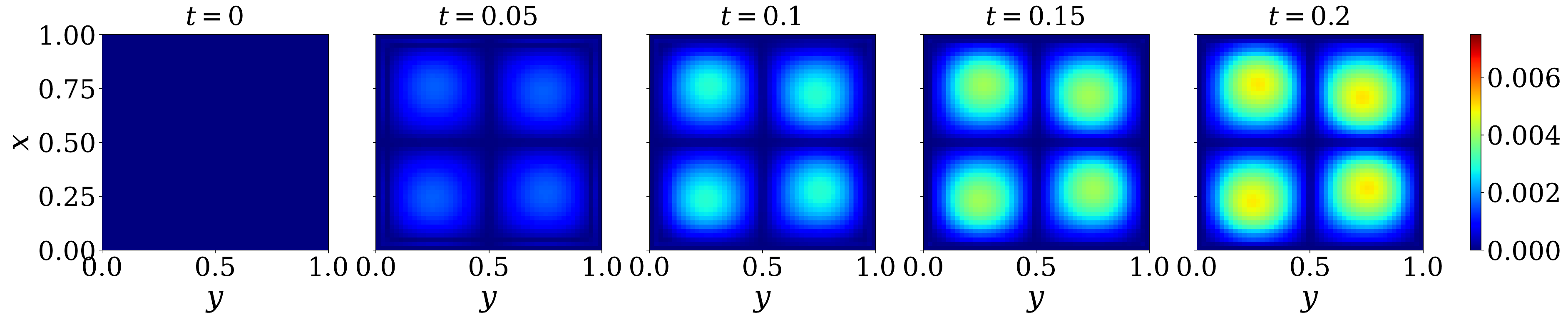}
        \caption{Std (S-SPH)}
    \end{subfigure}
    \begin{subfigure}[b]{\textwidth}
        \centering
        \includegraphics[width=\textwidth]{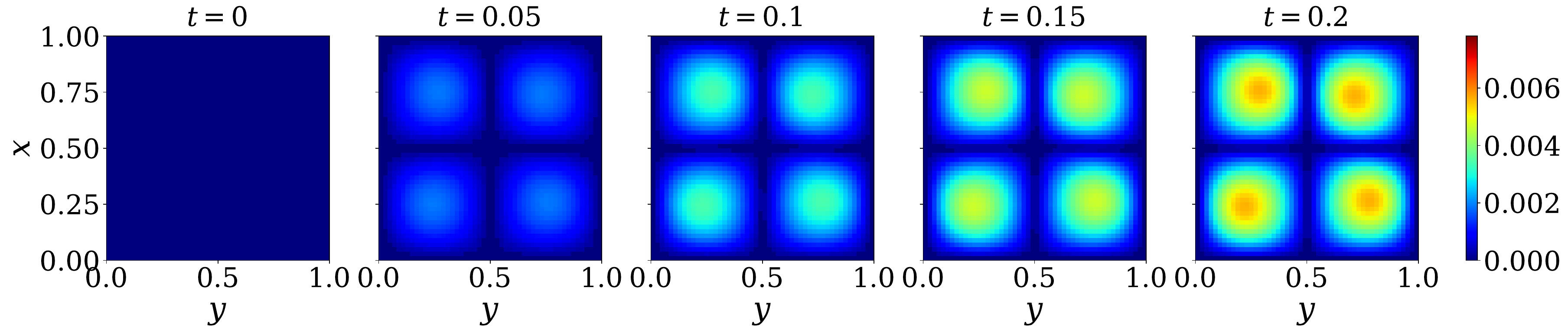}
        \caption{Std (MCS)}
    \end{subfigure}
    \caption{Time-evolving contour plots for Case~2 of the 2-D stochastic Burgers’ equation, where the viscosity is modeled as a spatially varying random field. 
    The first two rows show the mean velocity field obtained using S-SPH and MCS, respectively, while the bottom two rows present the corresponding standard deviation fields. 
    The figures illustrate the spatial pattern of uncertainty as the system evolves.
}
    \label{fig:ex3_case3}
\end{figure}

Figs.\ref{fig:ex3_case1} and~\ref{fig:ex3_case3} summarize the results obtained for the two stochastic configurations considered, namely Case~1 (random Fourier initial field) and Case~2 (spatially varying random viscosity field), respectively. In both cases, the spatio-temporal evolution of the mean and standard deviation of the velocity field predicted by the proposed S-SPH framework is compared against the corresponding MCS reference.
For Case~1, Fig.~\ref{fig:ex3_case1} shows the time-resolved contours of the mean and standard deviation of the velocity field at selected time instants. The mean fields obtained using S-SPH closely match those computed via MCS throughout the simulation horizon, accurately capturing the nonlinear advection-driven transport and subsequent diffusive smoothing of the initial random field. The standard deviation contours further demonstrate that S-SPH reliably reproduces the spatial distribution and temporal decay of uncertainty, with excellent agreement observed across the entire domain. This indicates that the proposed method effectively propagates uncertainty arising from complex, spatially correlated initial conditions in a two-dimensional nonlinear setting.

Case~2, shown in Fig.~\ref{fig:ex3_case3}, represents a more challenging scenario, as uncertainty enters the governing equations multiplicatively through a spatially varying random viscosity field. Despite this added complexity, the S-SPH predictions remain in close agreement with the MCS reference for both the mean and standard deviation responses. The mean velocity fields exhibit nearly indistinguishable spatio-temporal patterns, indicating that S-SPH accurately captures the impact of heterogeneous diffusion on the evolving flow structures. The corresponding standard deviation fields reveal consistent spatial patterns and magnitudes between S-SPH and MCS, demonstrating that the proposed approach can robustly handle uncertainty in model coefficients that directly influence the dissipative mechanisms of the system.

Across both cases, the strong agreement between S-SPH and MCS highlights the ability of the proposed framework to accurately resolve first- and second-order statistics in nonlinear, multi-dimensional stochastic partial differential equations. Importantly, these results are achieved with a significantly reduced computational cost compared to brute-force MCS, underscoring the computational efficiency of S-SPH while maintaining high fidelity in uncertainty quantification. The results further confirm that S-SPH remains stable and accurate in the presence of both additive uncertainty (random initial conditions) and multiplicative uncertainty (random viscosity fields), making it a viable and scalable alternative to Monte Carlo methods for stochastic fluid dynamics problems.

\section{Conclusion}\label{sec:conclusion}
This work introduced a stochastic extension of the smoothed particle hydrodynamics framework, termed Stochastic SPH (S-SPH), for the efficient and accurate propagation of uncertainty in nonlinear partial differential equations. By embedding orthogonal polynomials within the SPH formulation, the proposed approach enables a systematic separation of stochastic and spatial variability while preserving the mesh-free, Lagrangian characteristics of SPH. Uncertainty in system parameters, initial conditions, and spatially varying coefficients was represented using random variables and random fields discretized through Karhunen–Lo\`eve and Fourier expansions. The stochastic response is represented using orthogonal polynomials and SPH kernel functions. A Galerkin projection in stochastic space transformed the governing stochastic partial differential equations into a coupled deterministic system governing the evolution of the expansion coefficients, enabling structured and computationally efficient uncertainty propagation.

The performance of S-SPH was evaluated on a hierarchy of benchmark problems of increasing complexity, including wave advection, and one- and two-dimensional Burgers’ equations with both parametric and random-field uncertainties. Across all cases, S-SPH accurately reproduced Monte Carlo simulation (MCS) statistics for both mean and variance, while achieving nearly three orders of magnitude reduction in computational cost relative to direct sampling approaches. Minor discrepancies observed in selected cases were attributable primarily to finite-sample convergence limitations of MCS rather than deficiencies in the proposed method. Overall, the results establish S-SPH as a robust, accurate, and scalable framework for uncertainty quantification in nonlinear mechanics problems. Its ability to handle both parametric and spatially distributed uncertainties, combined with favorable computational scaling and full compatibility with mesh-free discretizations, positions S-SPH as a compelling alternative to traditional Monte Carlo-based methods for high-dimensional and strongly nonlinear systems.
Future work will focus on extending S-SPH to multiphysics systems, adaptive stochastic representations, and large-scale three-dimensional simulations

\section*{Acknowledgements}

\section*{Code availability}
Upon acceptance, all the source codes to reproduce the results in this study will be made available to the public on GitHub by the corresponding author.

\section*{Competing interests} 
The authors declare no competing interests.

\appendix
\section{Derivation of the Symmetric Gradient Formulation}

In classical Smoothed Particle Hydrodynamics (SPH), the gradient of a field variable \(u(\bm{x})\) at the location of particle \(i\) is commonly approximated by
\begin{equation}\label{eq:grad_classic}
\nabla u(\bm{x}_i) \approx -\sum_{j} u_j\,\nabla W_{ij}\,\frac{m_j}{\rho_j}\,,
\end{equation}
where \(u_j = u(\bm{x}_j)\) is the value of the field at the neighboring particle \(j\),  \(W_{ij}=W(\bm{x}_i-\bm{x}_j; h)\) is the kernel function (with smoothing length \(h\)), \(m_j\) and \(\rho_j\) denote the mass and density associated with the particle at \(\bm{x}_j\), and \(\nabla W_{ij}\) is the gradient of the kernel with respect to \(\bm{x}_i\)

A desirable property in many SPH formulations is to cast the operator in a symmetric (or pairwise antisymmetric) form. In our context, the symmetric form of the gradient approximation is given by
\begin{equation}\label{eq:grad_sym}
\nabla u(\bm{x}_i) \approx \sum_{j} \left(u_i-u_j\right)\,\nabla W_{ij}\,\frac{m_j}{\rho_j}\,.
\end{equation}

\subsection*{Step 1: Exploiting the Zero-Sum Property of the Kernel Gradient}

An important property of the SPH kernel is that its gradient satisfies the zero-sum condition. In particular, for a radially symmetric kernel and a complete set of neighboring particles, one has
\begin{equation}\label{eq:zero_sum}
\sum_{j} \nabla W_{ij}\,\frac{m_j}{\rho_j} \approx \mathbf{0}\,.
\end{equation}
This relation follows from the fact that the kernel’s derivative is antisymmetric with respect to particle position differences and ensures that the numerical approximation does not introduce spurious gradients in constant fields.

\subsection*{Step 2: Adding a Zero Term}

Given that Eq.~\eqref{eq:zero_sum} holds, we multiply the zero sum by the field value \(u_i\) at the location of particle \(i\) to obtain
\[
u_i \sum_{j} \nabla W_{ij}\,\frac{m_j}{\rho_j} \approx \mathbf{0}\,.
\]
Since this term is identically zero (within the approximation), it may be added to Eq.~\eqref{eq:grad_classic} without affecting the consistency of the formulation.

\subsection*{Step 3: Recasting the Classical Formulation}

Adding the zero term to Eq.~\eqref{eq:grad_classic} gives
\begin{align}\label{eq:add_zero}
-\sum_{j} u_j\,\nabla W_{ij}\,\frac{m_j}{\rho_j}
& = -\sum_{j} u_j\,\nabla W_{ij}\,\frac{m_j}{\rho_j} + u_i \sum_{j} \nabla W_{ij}\,\frac{m_j}{\rho_j} \nonumber \\
& = \sum_{j} \left[ u_i - u_j \right] \nabla W_{ij}\,\frac{m_j}{\rho_j}\,.
\end{align}
Thus, we obtain the symmetric gradient approximation expressed in Eq.~\eqref{eq:grad_sym}.

\section{Derivation of the S-SPH Governing Equation}

In this section, we summarize the derivation of the stochastic Smoothed Particle Hydrodynamics (S-SPH) governing equation, for the third numerical example presented in this work.

\subsection{Example 3: 2D Burgers’ Equation}
\subsubsection*{Step 1: Define the First-Order Derivative \( u^{(1)} \)}

The first-order derivative of the velocity component \( u \) with respect to \( x \) for each particle \( j \) is calculated as:
\[
u_j^{(1)} = \sum_{k} m_k \frac{u_k - u_j}{\rho_k} \frac{\partial W(x_j - x_k, y_j - y_k, h)}{\partial x_k}
\]
where:
\begin{itemize}
    \item \( u_j^{(1)} \) represents the first-order derivative of \( u \) at particle \( j \).
    \item \( m_k \) is the mass of particle \( k \).
    \item \( \rho_k \) is the density of particle \( k \).
    \item \( W(x_j - x_k, y_j - y_k, h) \) is the SPH kernel function.
\end{itemize}

\subsubsection*{Step 2: Apply SPH Again to Obtain the Second-Order Derivative (Viscous Term)}

Using \( u_j^{(1)} \) calculated in Step 1, we apply the SPH formulation again to compute the second-order derivative for the viscous term:
\[
\nu \Delta u_j = \nu \sum_{k} m_k \frac{u_k^{(1)} - u_j^{(1)}}{\rho_k} \frac{\partial W(x_j - x_k, y_j - y_k, h)}{\partial x_k}
\]

\subsubsection{Final Reformulated Equations }

The corrected SPH formulation for a particle \( j \) in the 2D Burgers' equation, including both the advective and viscous terms, is:
\\
   \begin{align*}
   \frac{\partial u_j}{\partial t} = &- \sum_{k} m_k \left[ u_j (u_k - u_j) + v_j (u_k - u_j) \right] 
   \frac{\partial W(x_j - x_k, y_j - y_k, h)}{\partial x_k} \\
   &+ \nu \sum_{k} m_k \frac{u_k^{(1)} - u_j^{(1)}}{\rho_k} 
   \frac{\partial W(x_j - x_k, y_j - y_k, h)}{\partial x_k}
   \end{align*}
   H:
   \[
   u_j^{(1)} = \sum_{k} m_k \frac{u_k - u_j}{\rho_k} 
   \frac{\partial W(x_j - x_k, y_j - y_k, h)}{\partial x_k}
   \]

   \begin{align*}
   \frac{\partial v_j}{\partial t} = &- \sum_{k} m_k \left[ u_j (v_k - v_j) + v_j (v_k - v_j) \right] 
   \frac{\partial W(x_j - x_k, y_j - y_k, h)}{\partial y_k} \\
   &+ \nu \sum_{k} m_k \frac{v_k^{(1)} - v_j^{(1)}}{\rho_k} 
   \frac{\partial W(x_j - x_k, y_j - y_k, h)}{\partial y_k}
   \end{align*}
   where:
   \[
   v_j^{(1)} = \sum_{k} m_k \frac{v_k - v_j}{\rho_k} 
   \frac{\partial W(x_j - x_k, y_j - y_k, h)}{\partial y_k}
   \]

where:
\begin{itemize}
    \item The first term represents the advective component.
    \item The second term represents the viscous component, computed by applying SPH twice to obtain the second-order derivative.
\end{itemize}

\subsubsection{Polynomial Chaos Decomposition (PCD) for the Reformulated Equations}

Using Polynomial Chaos Decomposition (PCD), the stochastic versions of the reformulated SPH equations for the \( u \) and \( v \) components are:

\begin{align*}
   \frac{\partial u_{l,j}}{\partial t} = &- \sum_{k} m_k \left[ u_{i,j} (u_{m,k} - u_{m,j}) + v_{i,j} (u_{m,k} - u_{m,j}) \right] 
   \frac{\partial W(x_j - x_k, y_j - y_k, h)}{\partial x_k} \, \langle \Phi_i(\omega) \Phi_m(\omega), \Phi_l(\omega) \rangle \\
   &+ \nu \sum_{k} m_k \frac{u_{m,k}^{(1)} - u_{m,j}^{(1)}}{\rho_k} 
   \frac{\partial W(x_j - x_k, y_j - y_k, h)}{\partial x_k} \, \langle \Phi_m(\omega), \Phi_l(\omega) \rangle
   \end{align*}
   where:
   \[
   u_{m,j}^{(1)} = \sum_{k} m_k \frac{u_{m,k} - u_{m,j}}{\rho_k} 
   \frac{\partial W(x_j - x_k, y_j - y_k, h)}{\partial x_k}
   \]

\begin{align*}
   \frac{\partial v_{l,j}}{\partial t} = &- \sum_{k} m_k \left[ u_{i,j} (v_{m,k} - v_{m,j}) + v_{i,j} (v_{m,k} - v_{m,j}) \right] 
   \frac{\partial W(x_j - x_k, y_j - y_k, h)}{\partial y_k} \, \langle \Phi_i(\omega) \Phi_m(\omega), \Phi_l(\omega) \rangle \\
   &+ \nu \sum_{k} m_k \frac{v_{m,k}^{(1)} - v_{m,j}^{(1)}}{\rho_k} 
   \frac{\partial W(x_j - x_k, y_j - y_k, h)}{\partial y_k} \, \langle \Phi_m(\omega), \Phi_l(\omega) \rangle
   \end{align*}
   where:
   \[
   v_{m,j}^{(1)} = \sum_{k} m_k \frac{v_{m,k} - v_{m,j}}{\rho_k} 
   \frac{\partial W(x_j - x_k, y_j - y_k, h)}{\partial y_k}
   \]

In these PCD expansions:
\begin{enumerate}
  
  \item  \( u_{l,j} \) and \( v_{l,j} \) represent the PCD-based coefficients for each velocity component.
  \item \( \Phi_i(\omega) \), \( \Phi_m(\omega) \), and \( \Phi_l(\omega) \) are the polynomial basis functions for the PCD.
  \item \( \langle \cdot , \cdot \rangle \) denotes the inner product used to project the random fields onto the polynomial basis.

\end{enumerate}

 We can approximate the number of bases required for this case. Since we have a total of \(q\) = 366 independent random variables, and even if we consider a polynomial basis for each of the random variables of order up to \(P\) = 2, from equation 11, we will get 65,768 basis functions in the expansion of the random field \(u\). This will create a problem of high dimensionality, and it'll be very difficult to solve it using normal computational resources. Hence, we will use KL expansion to reduce the dimensionality of the problem and make it easily solvable without taking up much computational resources.


\end{document}